\documentstyle[12pt,epsfig]{article}

\begin{document}
\def\cmm{cm$^{-2}$}
\def\cmmm{cm$^{-3}$}
\def\kms{km~s$^{-1}$}
\def\ho{km~s$^{-1}$ Mpc$^{-1}$}
\def\mf{$\times 10^{-5}$}
\def\mt{$\times 10^{-10}$}
\def\Lya{Ly$\alpha$}
\def\lya{Ly$\alpha$}
\def\lyaf{Ly$\alpha$ forest}
\def\Lyb{Ly$\beta$}
\def\Lyg{Ly$\beta$}
\def\nhi{$N_{HI}$}
\def\nhe{$N_{HeII}$}
\def\ETA{$\eta$}
\def\ob{$\Omega _b$}
\def\om{$\Omega _m$}
\def\hsv{$h_{70}^{-2}$}
\def\d0014 {Q0014+8118}
\def\d0130 {Q0130-4021}
\def\d0827 {Q08279+5255}
\def\d1009 {Q1009+2956}
\def\d1251 {Q1251+3644}
\def\d1718 {Q1718+4807}
\def\d1759 {Q1759+7539}
\def\d1937 {Q1937-1009}
\def\yp{$Y_p$}
\def\het{$^3$He}
\def\hef{$^4$He}
\def\lisx{$^6$Li}
\def\lisv{$^7$Li}

\title{Review of Big Bang Nucleosynthesis and Primordial Abundances}

\author{David Tytler, John M. O'Meara, Nao Suzuki \& Dan Lubin\\
Center for Astrophysics and Space Sciences;
\\ University of California, San Diego; \\ MS 0424; La Jolla; CA
92093--0424\\}

\maketitle

\begin{abstract}

Big Bang Nucleosynthesis (BBN) is the synthesis 
of the light nuclei, Deuterium, \het, \hef ~ and \lisv ~ during the
first few minutes of the universe.
This review concentrates on recent improvements in the measurement of the
primordial (after BBN, and prior to modification) abundances of these nuclei.
We mention improvement in the standard theory, and the non-standard extensions
which are limited by the data.

We have achieved an order of magnitude improvement in the
precision of the measurement of primordial D/H, using
the HIRES spectrograph on the W. M. Keck telescope to measure
D in gas with very nearly primordial abundances towards quasars.  From 
1994 -- 1996, it appeared that there could be a factor of ten range in
primordial D/H, but today four examples of low D are secure. High D/H should be
much easier to detect, and since there are no convincing examples, it must
be extremely rare or non-existent.
All data are consistent with a single low value for D/H, and the
examples which are consistent with high D/H are readily interpreted as
H contamination near the position of D.

The new D/H measurements give the most accurate value for the baryon to
photon ratio, \ETA , and hence the cosmological baryon density.
A similar density is required to explain the amount of \Lya\ absorption
from neutral Hydrogen in the intergalactic medium (IGM)
at redshift $z \simeq 3$, and
to explain the fraction of baryons in local clusters of galaxies.

The D/H measurements
lead to predictions for the abundances of the other light nuclei, which
generally agree with measurements.

The remaining differences with some measurements can be explained by
a combination of measurement and analysis errors or changes in the
abundances after BBN. The measurements do not
require physics beyond the standard BBN model.
Instead, the agreement between the abundances is used to limit the
non-standard physics.

New measurements are giving improved understanding of the
difficulties in estimating the abundances of all the light nuclei,
but unfortunately in most cases we are not yet seeing much improvement in the
accuracy of the primordial abundances.
Since we are now interested in the highest accuracy and reliability for all
nuclei, the few objects with the most extensive observations give by far the
most convincing results.

Earlier measurements of \hef\ may have obtained too low a value because
the He emission line strengths were reduced by undetected stellar
absorption lines.
The systematic errors associated with the \hef\ abundance have frequently
been underestimated in the past, and this problem persists.
When two groups use the same data and different ways to estimate the electron
density and \hef\ abundance, the results differ by more than the
quoted systematic errors. While the methods used by Izotov \& Thuan
\cite{hefdata} seem to be an advance on those used before, the other method is reasonable,
and hence the systematic error should encompass the range in results.

The abundance of \lisv\ is measured
to high accuracy, but we do not know how much was produced prior to the
formation of the stars, and how much was destroyed (depleted) in the stars.
\lisx\ helps limit the amount of depletion of \lisv , but by
an uncertain amount since it too has been depleted.

BBN is successful because it
uses known physics and measured cross-sections for the nuclear reactions.
It gives accurate predictions for the abundances of five light nuclei
as a function of the one free parameter \ETA .
The other initial conditions seem natural: the universe began
homogeneous and hotter than $T> 10^{11}$ K (30 Mev).
The predicted abundances agree with most observations, and
the required \ETA\ is consistent with other,  less accurate, measurements
of the baryon density.

New measurements of the baryon
density, from the CMB, clusters of galaxies and the \lya\ forest, will give
\ETA . Although the accuracy might not exceed
that obtained from D/H, this is an important advance because
BBN then gives abundance predictions with no adjustable parameters.

New measurement in the coming years will give improved accuracy.
Measurement of D/H in many more quasar spectra would improve the accuracy of
D/H by a factor of a few, to a few percent, but even with improved
methods of selecting the target quasars, this would need much more
time on the largest telescopes.
More reliable \hef\ abundances might be obtained from spectra which have
higher spectral and spatial resolution, to help
correct for stellar absorption, higher signal to noise to show weaker
emission lines, and more galaxies with low metal abundances, to minimize
the extrapolation to primordial abundances.
Measurements of \lisx , Be and Boron in the same stars
and observations of a variety of stars should give improved models for
the depletion of \lisv\ in halo stars, and hence tighter constraints on
the primordial abundance.
However, in general, it is hard to think of any new methods which could give
any primordial abundances with an order of magnitude higher accuracy than
those used today.
This is a major unexploited opportunity, because it means
that we can not yet test BBN to the accuracy of the predictions.
\end{abstract}

\section{Introduction}

There are now four main observations which validate the Big Bang theory:
the expansion of the universe,
the Planck spectrum of the Cosmic Microwave Background (CMB),
the density fluctuations seen in the slight CMB anisotropy and in the local
galaxy distribution, and BBN. Together, they show that the universe began
hot and dense \cite{turn99}.

BBN occurs at the earliest times at which we have a
detailed understanding of physical processes. It makes
predictions which are relatively precise (10\% -- 0.1\%), and which have been
verified with a variety of data.
It is critically important that the standard theory (SBBN) predicts the
abundances of several light nuclei (H, D, \het\, \hef\, and \lisv\ ) as
a function of a single cosmological
parameter, the baryon to photon
ratio, $\eta \equiv n_b/n_\gamma$ \cite{kt90}.  The ratio of any two
primordial abundances should give $\eta$, and
the measurement of the other three tests the theory.

The abundances of all the light elements have been measured
in a number of terrestrial and astrophysical environments.
Although it has often been hard to decide when these abundances
are close to primordial, it has been clear for decades
(e.g. \cite{ree73}, \cite{wag73})
that there is general agreement with the BBN predictions for all the
light nuclei.
The main development in recent years has been the increased accuracy of
measurement.
In 1995 a factor of three range in the baryon density was considered
$\Omega _b = 0.007 - 0.024$. The low end of this range allowed no
significant dark baryonic matter. Now the new D/H measurements towards
quasars give
$\Omega _b = 0.019 \pm 0.0024$ (95\%) -- a 13\% error, and there have been
improved measurements of the other nuclei.

\subsection{Other Reviews}

Many reviews of BBN have been published recently: 
e.g. \cite{bur99b}, ~\cite{oli99b}, ~\cite{oli99c},
~\cite{sar99}, ~\cite{sch98}, ~\cite{sch98a}, ~\cite{ste98} and 
~\cite{cst95b}, some of which are lengthy: e.g.
\cite{oli99a}, ~\cite{sar96} and \cite{boes85}.
All modern cosmology texts contain a summary.
Several recent books contain the proceedings of meetings on this
topic:  \cite{cra95}, \cite{holt96}, \cite{prab98} and
\cite{prag98}. The 1999 meeting of the
International Astronomical Union (Symposium 198 in Natal, Brazil) was on
this topic, as are many reviews in
upcoming special volumes of Physics Reports and New Astronomy, both in honor of
the major contributions by David N. Schramm.

\section{Physics of BBN}

Excellent summaries are given in most books on cosmology
e.g.: \cite{boern93}, \cite{cos text},
\cite{kt90}, \cite{pea99}, \cite{prad93},
and most of the reviews listed above, including
\cite{bern91}, and
\cite{sch98a}.

\subsection{Historical Development}

The historical development of BBN is reviewed by
\cite{het93}, \cite{cst95b},
\cite{kra96}, \cite{sch98} and \cite{bur99b}.
Here we mention a few of the main events.

The search for the origin of the elements lead to
the modern Big Bang theory in the early 1950s.
The expansion of the universe was widely accepted when
Lemaitre \cite{lem31} suggested that the universe began in an explosion of a
dense unstable ``primeval atom".
By 1938 it was well established that the abundances of the
elements were similar in different astronomical locations, and hence
potentially of cosmological significance.
Gamow \cite{gamow42}, \cite{gamow46} asked whether nuclear reactions in the early universe
might explain the
abundances of the elements. This was the first examination of the
physics of a dense expanding early universe, beyond the mathematical
description of general relativity, and over
the next few years this work developed into the modern big bang theory.
Early models started with pure neutrons, and gave final abundances
which depended on the unknown the density during BBN. 
Fermi \& Turkevich showed that the lack of stable nuclei with mass 5 and 8
prevents significant production of nuclei more massive than \lisv,
leaving \hef\ as the most abundant nucleus after H.
Starting instead with all possible species,
Hayashi \cite{hay50} first calculated the neutron to proton (n/p) ratio during
BBN, and
Alpher \cite{alp48} realized that radiation would dominate
the expansion.  By 1953 \cite{afh53} the basic physics of
BBN was in place.
This work lead directly to the prediction of the CMB (e.g. Olive 1999b~\cite{oli99b}),
it explained the origin of D, and gave
abundance predictions for \hef\ similar to those obtained
today with more accurate cross-sections.

The predicted abundances have changed little in recent years, following
earlier work by Peebles (1964) \cite{pee66},                                    
Hoyle \& Tayler (1964) \cite{hoy64}, and
Wagoner, Fowler \& Hoyle (1967) \cite{wag67}. The accuracy of the               
theory calculations have been improving, and they remain much more              
accurate than the measurements.                                                 
For example, the fraction of the mass of all baryons which is \hef , \yp ,
is predicted to within $\delta$\yp\ $< \pm 0.0002$  \cite{lop99}.
In a recent update, Burles et al. \cite{bur99b} uses Monte-Carlo realizations
of reaction rates to find that the previous estimates of
the uncertainties in the abundances for a given \ETA\ were a factor of
two too large.

\section{Key Physical Processes}

\subsection{Baryogenesis}

The baryon to photon ratio \ETA\ is determined during baryogenesis
\cite{kt90}, \cite{rio98}, \cite{rio99}.
It is not known when baryogenesis occurred.
Sakharov \cite{sak67} noted that three conditions are required:
different interactions for matter and anti-matter (CP violation),
interactions which change the baryon number, and
departure from thermodynamic equilibrium.
This last condition may be satisfied in a first order phase transition,
the GUT transition at $10^{-35}$ s,
or perhaps the electroweak transition at $10^{-11}$ s.
If baryogenesis occurred at the electroweak scale, then future measurements
may lead to predictions for \ETA , but if,
alternatively, baryogenesis is at the GUT or inflation scale,
it will be very hard to predict \ETA\ (J. Ellis personal communication).

The matter/anti-matter asymmetry of the universe (the \ETA\ value)
is attracting discussion in the popular science press
because of the inauguration of major experiments to study CP violation in
B mesons (\cite{irion99}, \cite{qui98}; Economist, May 8 1999, 85-87).

\subsection{The main physical processes in BBN}

At early times, weak reactions keep the n/p ratio close to the
equilibrium Boltzmann ratio.
As the temperature, T, drops, n/p decreases.
The n/p ratio is fixed (``frozen in") at a value of about 1/6
after the weak reaction rate is slower
than the expansion rate. This is at about 1 second, when $T \simeq 1$MeV.
The starting reaction n+p $^{ \rightarrow} _{ \leftarrow } D + \gamma$
makes D.
At that time photodissociation of D is rapid because of the high entropy
(low \ETA ) and this prevents significant
abundances of nuclei until, at 100 sec., the temperature has dropped
to 0.1 MeV, well below the binding energies of the light nuclei.
About 20\% of free neutrons decay prior to being incorporated into nuclei.
The \hef\ abundance is then given approximately by assuming that all remaining
neutrons are incorporated into \hef .

\begin{figure}
\centerline{\psfig{file=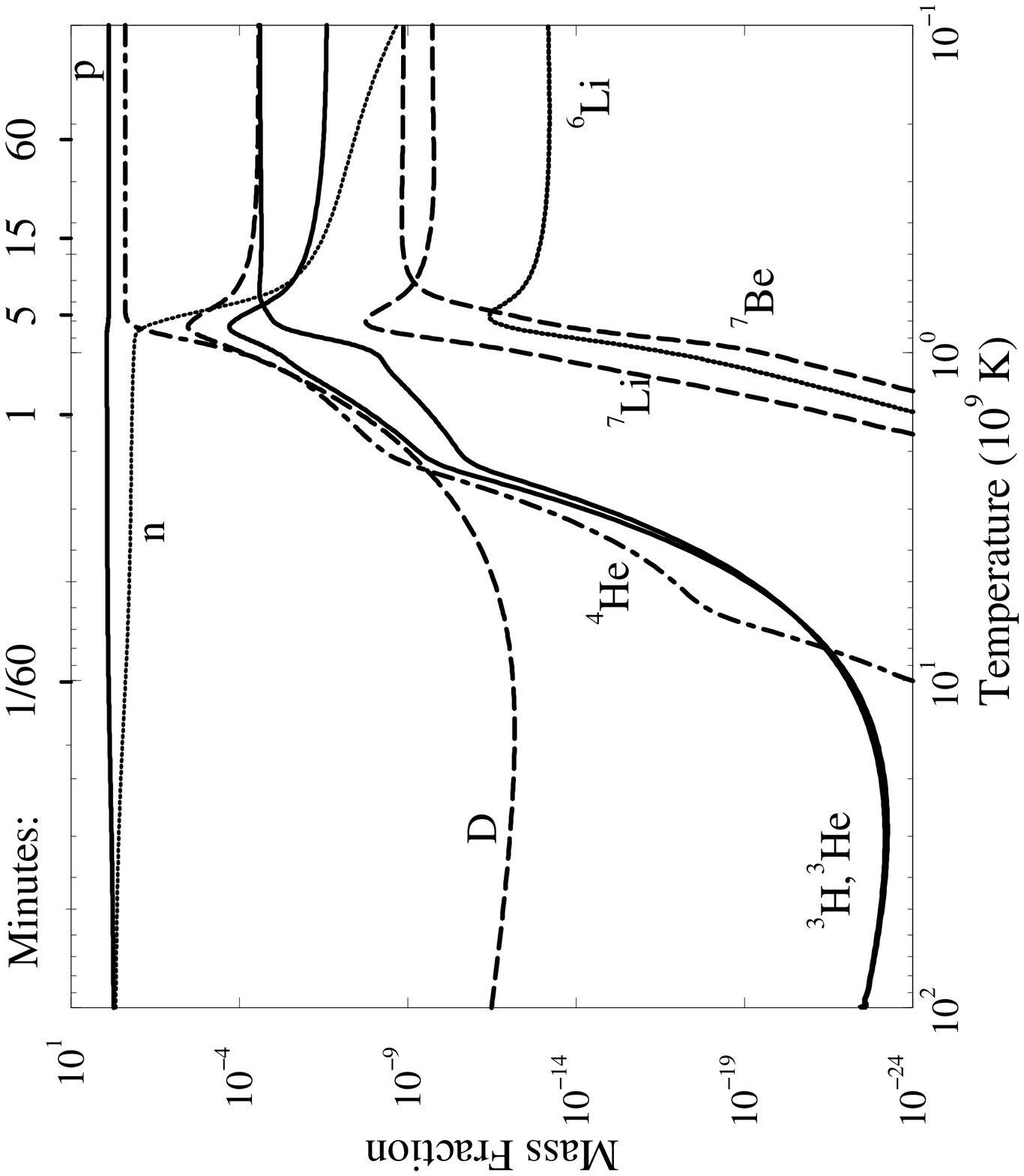, height=6.0in}}
\caption{ Mass fraction of nuclei as a function of temperature
for $\eta = 5.1 \times 10^{-10}$, from Nollet \& Burles (1999) and
Burles et al. (1999).}

\label{evolution bw}
\end{figure}

\begin{figure}
\centerline{\psfig{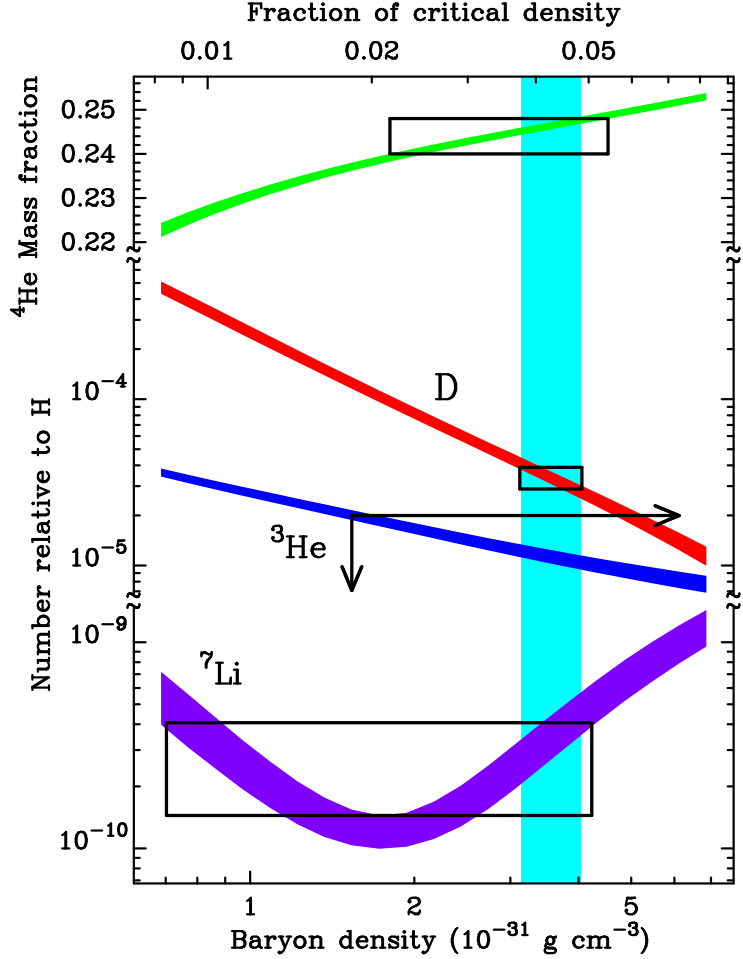}}
\caption{Abundances expected for the light nuclei $^4$He, D, $^3$He and
$^7$Li (top to bottom) calculated in standard BBN. New estimates of the
nuclear cross-section errors from Burles et al. (1999a) and Nollet \&
Burles (1999) were used to
estimate the 95\% confidence intervals which are shown by the vertical widths
of the abundance predictions.
The horizontal scale, $\eta$, is the one free parameter in the calculations.
It is expressed in units of the baryon density or critical density for a Hubble
constant of 65 kms$^{-1}$Mpc$^{-1}$.
The 95\% confidence intervals for data,
shown by the rectangles, are from Izotov and Thuan 1998a ($^4$He);
Burles \& Tytler 1998a (D);
Gloeckler \& Geiss 1996 ($^3$He);
Bonifacio and Molaro 1997 ($^7$Li extended upwards by a factor of two to
allow for possible depletion).}
\label{all abundances}
\end{figure}

The change in the abundances over time for one \ETA\ value is shown in Figure
\ref{evolution bw}, while the dependence of the final
abundances on \ETA\ is shown in Figure \ref{all abundances}, together with
some recent measurements.

In general, abundances are given by two cosmological parameters,
the expansion rate and \ETA . Comparison with the strength of the
weak reactions gives the n/p ratio, which determines \yp .
\yp\ is relatively independent of \ETA\ because n/p depends on
weak reactions between nucleons and leptons (not pairs of nucleons),
and temperature.
If \ETA\ is larger, nucleosynthesis starts earlier, more nucleons
end up in \hef , and \yp\ increases slightly. D and \het\
decrease simultaneously in compensation.
Two channels contribute to the abundance of \lisv\ in the \ETA\ range of
interest, giving the same \lisv\ for two values of \ETA .

\section{Measurement of Primordial Abundances}

The goal is to measure the primordial abundance ratios of the light
nuclei made in BBN. We normally measure the ratios of the abundances
of two nuclei in the same gas, one of which is typically H,
because it is the easiest to measure.

The two main difficulties are the accuracy of the measurement and
departures from primordial abundances.
The state of the art today (1 $\sigma$) is about 3\% for \yp , 10\% for
D/H and 8\% for \lisv , for each object observed.
These are random errors. The systematic errors are hard to estimate,
usually unreliable, and potentially much larger.

By the earliest time at which we can observe objects, redshifts $z \simeq 6$,
we find heavy elements from stars in most gas. Although we expect that
large volumes of the intergalactic medium (IGM) remain primordial today
\cite{gne97}, we do not know how to obtain accurate abundances
in this gas. Hence we must consider possible modifications of abundances.
This is best done in gas with the lowest abundances of heavy elements,
since this gas should have the least deviations caused by stars.

The nuclei D, \het, \lisx\ and \lisv\ are  all fragile and readily burned
inside stars at relatively low temperatures of a few $10^6$ K.
They may appear depleted in the atmosphere of a star because the gas in
the star has been above the critical temperature, and they will be depleted in
the gas returned to the interstellar medium (ISM).
Nuclei \het , \lisv\ and especially \hef\ are also made in stars.

\subsection{From Observed to Primordial Abundances}

Even when heavy element abundances are low, it is difficult to prove that
abundances are primordial. Arguments include the following.

{\bf Helium} is observed in the ionized gas surrounding luminous young stars
(H~II regions), where O abundances are 0.02 to 0.2 times those in the sun.
The \hef\ mass fraction $Y$ in different galaxies
is plotted as a function of the abundance of O or N.
The small change in $Y$ with O or N is the clearest evidence that the $Y$
is almost entirely primordial (e.g. \cite{oli99b} Fig 2).
Regression gives the predicted \yp\ for zero O or N \cite{he regress}.
The extrapolation is a small extension beyond the observed range,
and the deduced primordial \yp\ is within the range of
$Y$ values for individual H~II regions.
The extrapolation should be robust
\cite{yp regress}, but some algorithms are
sensitive to the few galaxies
with the lowest metal abundances, which is dangerous because at least one
of these values was underestimated by Olive, Skillman \& Steigman \cite{osw99 yp}.

For {\bf deuterium} we use a similar argument. The observations are made in
gas with two distinct metal abundances.
The quasar absorbers have from 0.01 to 0.001 of the solar C/H,
while the ISM and pre-solar observations are near solar.
Since D/H towards quasars is twice that in the ISM,
50\% of the D is destroyed when abundances rise to near the solar level, and
less than 1\% of D is expected to be destroyed in the quasar absorbers,
much less than the random errors in individual measurements of D/H.
Since there are no other known processes which destroy or make significant D
(e.g.\cite{ree73}, \cite{inhom and high obb}),
we should be observing primordial D/H in the quasar absorbers.

{\bf Lithium} is more problematic.
Stars with a variety of low heavy element abundances
(0.03 -- 0.0003 of solar) show
very similar abundances of \lisv\ (\cite{li sys} Fig 3), which should be
close to the primordial value.  Some use the observed values in these ``Spite plateau"
stars as the BBN abundance, because of the small scatter and lack of variation with
the abundances of other elements, but three factors should be considered.  First,
the detection of \lisx\ in two of these stars suggests that
both \lisx\ and some \lisv\ was been created prior to the formation of these
stars.  Second, the possible increase in the abundance of \lisv\ with the iron
abundance also indicates that the \lisv\ of the plateau stars is not
primordial.
If both the iron and the enhancement in the \lisv\ have the same origin
we could extrapolate back to zero metals \cite{li summary}, as for \hef , but
the enhanced \lisv\ may come from cosmic ray interactions in the ISM,
which makes extrapolation less reliable.
Third, the amount of depletion is hard to estimate. Rotationally induced
mixing has a small effect because there is little scatter on the
Spite plateau, but other mechanisms may have depleted \lisv .
In particular, gravitational settling should have occurred, and
left less \lisv\ in the hotter plateau stars, but this is not seen,
and we do not know why. More on this later.

The primordial abundance of \het\ is the hardest to estimate, because
stars are expected to both make and destroy this isotope, and there are no
measurements in gas with abundances well below the solar value.

\subsection{Key observational Requirements}

By way of introduction to the data, we list some of the key goals of ongoing
measurements of the primordial abundances.

\begin{itemize}
\item \hef : High accuracy, robust measurement in a few places
with the lowest metal abundances.
\item \het : Measurement in gas with much lower metal abundances, or
an understanding of stellar production and destruction and the results
of all stars integrated over the history of the Galaxy
(Galactic chemical evolution).
\item D: The discovery of more quasar absorption systems with minimal H
contamination.
\item \lisv : Observations which determine the amount of depletion in
halo stars, or which avoid this problem. Measurement of \lisx , Be and B
to help estimate production prior to halo star formation, and
subsequent depletion.
\end{itemize}

Since we are now obtaining ``precision" measurements, it now seems best
to make a few measurements with the highest possible accuracy and
controls, in places with the least stellar processing, rather than
multiple measurements of lower accuracy.
We will now discuss observations of each of the nuclei, and especially
D, in more detail.

\section{Deuterium in quasar spectra}

The abundance of deuterium (D or $^2$H) is the most sensitive measure of
the baryon density \cite{wag73}.
No known processes make significant D, because it is so fragile
(\cite{ree73}, \cite{eps76}, \cite{boy89} and
\cite{destruct of d}).
Gas ejected by stars should contain zero D, but substantial H, thus
D/H decreases over time as more stars evolve and die.

We can measure the primordial abundance in quasar spectra. The measurement is
direct and accurate, and with one exception, simple.
The complication is that the absorption by D is often contaminated or
completely obscured by the absorption from H, and even in the rare
cases when contamination is small, superb spectra are required to
distinguish D from H.

Prior to the first detection of D in quasar spectra
\cite{tyt96b}, D/H was measured in the
ISM and the solar system.  The primordial abundance is
larger, because D has been destroyed in stars. Though generally
considered a factor of a few, some papers considered a factor of ten
destruction \cite{aud86}.
At that time, most measurements of \hef\ gave low abundances, which
predict a high primordial D/H, which would need to be depleted
by a large factor to reach ISM values \cite{vid84}.

Reeves, Audouze, Fowler \& Schramm \cite{ree73} noted that the measurement of
primordial D/H
could provide an excellent estimate of the cosmological baryon density,
and they used the ISM \het\ +D to concluded, with great caution,
that primordial D/H was plausibly $7 \pm 3$ \mf .

Adams \cite{adams76} suggested that it
might be possible to measure primordial D/H towards low metallicity
absorption line systems in the spectra of high redshift quasars.  This gas is
in the outer regions of galaxies or in the IGM, and it
is not connected to the quasars.
The importance of such measurements was well known in the field since late
1970s \cite{web91}, but
the task proved too difficult for 4-m class telescopes (\cite{chaf85},
\cite{chaf86}, \cite{cars94}). The high SNR QSO spectra obtained with
the HIRES echelle spectrograph \cite{vog94} on the W.M. Keck 10-m telescope
provided the breakthrough.

There are now three known absorption systems in which D/H is low:
first, D/H $= 3.24 \pm 0.3 \times 10^{-5}$ in the $z_{abs} = 3.572$ Lyman limit
absorption system (LLS) towards quasar 1937--1009 \cite{tyt96b}, \cite{bur98b};
second, D/H = $4.0 ^{+0.8}_{-0.6} \times 10^{-5}$ in the $z_{abs} = 2.504$
LLS towards quasar 1009+2956 \cite{bur98a}, and third, D/H $< 6.7$ \mf\ towards
quasar 0130--4021 \cite{kirk 0130}. This last case is the simplest
found yet, and seems especially secure because the entire Lyman
series is well fit by a single velocity component.  The velocity of
this component and its column density are well determined because
many of its Lyman lines are unsaturated. Its \lya\ line is simple
and symmetric, and can be fit using the H parameters determined by
the other Lyman series lines, with no additional adjustments for
the \lya\ absorption line.  There is barely enough absorption at
the expected position of D to allow low values of D/H, and there
appears to be no possibility of high D/H. Indeed, the spectra of
all three QSOs are inconsistent with high D/H.

There remains uncertainty over a case at $z_{abs} = 0.701$
towards quasar 1718+4807, because we lack spectra of the Lyman series
lines which are needed to determine the velocity distribution of the
Hydrogen, and these spectra are of unusually low signal to noise,
with about 200 times fewer photons per kms$^{-1}$ than those from Keck.
Webb et al. \cite{web97a}, \cite{web97b} assumed a single hydrogen
component and found D/H = $25 \pm 5 \times 10^{-5}$, the
best case for high D/H. Levshakov et
al. \cite{1009} allow for non-Gaussian velocities and find D/H $\sim 4.4
\times 10^{-5}$, while Tytler et al. \cite{tyt98a} find $8 \times 10^{-5} <$
D/H $<57 \times 10^{-5}$ (95\%) for a single Gaussian component, or
D/H as low as zero if there are two hydrogen components, which is not
unlikely. This quasar is then also consistent with low D/H.

Recently Molaro et al. \cite{molaro99} claimed that D/H might be low in
an absorber
at $z=3.514$ towards quasar APM 08279+5255, though they noted that
higher D/H was also possible. Only one H~I line, \lya , was used to
estimate the hydrogen column density
\nhi\ and we know that in such cases the column density
can be highly uncertain. Their Figure 1 (panels a and b) shows that
there is a tiny difference between D/H = 1.5 \mf\ and 21 \mf , and it
is clear that much lower D is also acceptable because there can be H
additional contamination in the D region of the spectrum.
Levshakov et al. \cite{08279} show that \nhi = 15.7 (too low to show D)
gives an excellent fit to
these spectra, and they argue that this is a more realistic result
because the metal abundances and temperatures are then normal, rather than
being anomalously low with the high \nhi\ preferred by Molaro et al.

The first to publish a D/H estimate using high signal to noise
spectra from the Keck telescope with the HIRES spectrograph were
Songaila et al. \cite{song94}, who reported an upper
limit of D/H $< 25 \times 10^{-5}$ in the $z_{abs} = 3.32$ Lyman limit
system (LLS) towards quasar 0014+813.  Using different spectra, Carswell
et al. \cite{cars94} reported $<60 \times 10^{-5}$ in the same object, and
they found no reason to think that the deuterium abundance might be as
high as their limit. 
Improved spectra
\cite{bur99c} support the early conclusions: D/H $< 35 \times
10^{-5}$ for this quasar. High D/H is allowed, but is highly
unlikely because the
absorption near D is at the wrong velocity, by $17 \pm 2$ \kms,
it is too wide, and it does not have the expected distribution of absorption
in velocity,
which is given by the H absorption.
Instead this absorption is readily explained
entirely by H (D/H $\simeq 0$) at a different redshift.

Very few LLS have a velocity structure simple enough to show
deuterium. Absorption by H usually absorbs most of the quasar flux near
where the D line is expected, and hence we obtain no
information of the column density of D. In these extremely common cases,
very high D/H is allowed, but only because we have essentially no
information.

All quasar spectra are consistent with
low primordial D/H ratio, D/H $\sim 3.4 \times 10^{-5}$.
Two quasars (1937--1009 \& 1009+2956)
are inconsistent with D/H $\ge 5 \times 10^{-5}$, and the third
(0130--4021)  is inconsistent with D/H $\ge 6.7 \times 10^{-5}$.
Hence D/H is low in these three places.
Several quasars allow high D/H, but in all cases this can be explained by
contamination by H, which we discuss more below, because this is the key
topic of controversy.

\subsection{ISM D/H}

Observations of D in the ISM are reviewed by Lemoine et al. \cite{lem99}.
The first measurement in the ISM, D/H $=1.4 \pm 0.2$ \mf , using Lyman
absorption lines observed with the Copernicus satellite \cite{rog73},
have been confirmed with superior HST spectra.
A major program by Linsky et al. \cite{lin93}, \cite{lin95} has given a 
secure value for local ISM ($<20$ pc) D/H = $1.6 \pm 0.1$ \mf .

Some measurements have indicated variation, and especially
low D/H, in the local and more distant ISM towards a few stars
\cite{vid84}, \cite{lem99}. Vidal-Madjar \& Gry \cite{vid84}
concluded that the different lines of sight gave different
D/H, but those early data may have been inadequate to
quantify complex velocity structure \cite{mcc92}.
Variation is expected, but at a low level, from different amounts of
stellar processing and infall of IGM gas, which leaves differing D/H
if the gas is not mixed in a large volume.

Lemoine et al. \cite{d n 191b2b} suggested variation of D/H towards
G191-B2B, while Vidal-Madjar et al. \cite{vid98} described the variation as real,
however new STIS spectra do
not confirm this, and give the usual D/H value.
The STIS spectra \cite{sah} show a simpler velocity
structure, and a lower flux at the D velocity, perhaps because of difficulties
with the background subtraction in the GHRS spectra.

H{$\rm \acute e$}brand et al. \cite{heb99} report the possibility of low D/H $< 1.6$
\mf\ towards Sirius A, B.

The only other instance of unusually low D/H from recent data is
D/H $= 0.74^{+0.19}_{-0.13}$ \mf ~(90\%) towards the star $\delta$ Ori
\cite{jen99}. We would much like to see improved data on this
star, because a new instrument was used, the signal to noise is
very low, and the velocity distribution of the
D had to be taken from the N~I line, rather than from the H~I.

Possible variations in D/H in the local ISM have no obvious connections to
the D/H towards quasars, where the absorbing clouds are 100 times larger,
in the outer halos young of galaxies rather in the dense disk,
and the influence of stars should be slight because heavy element abundances
are 100 to 1000 times smaller.

Chengalur, Braun \& Burton \cite{chen97} report D/H $=3.9 \pm 1.0$ \mf\ from
the marginal detection of
radio emission from the hyper-fine transition of D at 327 MHz  (92 cm).
This observation was of the ISM in the direction of the Galactic anti-center,
where the molecular column density is low, so that most D should be atomic.
The D/H is higher than in the local ISM, and similar to the primordial value,
as expected, because there has been little stellar processing in this direction.

Deuterium has been detected in molecules in the ISM. Some of
these results are considered less secure because of fractionation and in
low density regions, HD is more readily destroyed by ultraviolet radiation,
because its abundance is too low to provide self shielding,
making HD/H$_2$ smaller than D/H.

However, Wright et al. \cite{writ99} deduce D/H = $1.0 \pm 0.3 \times 10^{-5}$
from the first detection of the 112 $\mu $m pure rotation line of HD
outside the solar system, towards the dense warm molecular clouds in
the Orion bar, where most D is expected to be in HD, so that
D/H $\simeq$ HD/H$_2$.
This D/H is low, but not significantly lower than in the local ISM, especially
because the H$_2$ column density was hard to measure.

Lubowich et al. \cite{lub98a}, \cite{lub98b} report D/H = $0.2 \pm 1$ \mf\ 
from DCN in the Sgr A molecular cloud near the Galactic center,
later revised to 0.3 \mf\  (private communication 1999).
This detection has two important implications.
First, there must be a source of D, because all of the gas here
should have been inside at least one star, leaving no detectable D.
Nucleosynthesis is ruled out because this would
enhance the Li and B abundances by orders of magnitude,
contrary to observations. Infall of less processed gas seems likely.
Second, the low D/H in the Galactic center implies that there
is no major source of D, otherwise D/H could be
very high. However, this is not completely secure,
since we could imagine a fortuitous cancellation between creation
and destruction of D.

We eagerly anticipate a dramatic improvement in the data on the ISM
in the coming years. The FUSE satellite, launched in 1999, will measure
the D and H Lyman lines towards thousands of stars and a few quasars,
while SOFIA (2002) and FIRST (2007) will measure HD in dense molecular clouds.
The new GMAT radio telescope should allow secure detection of D 82 cm
emission from the outer Galaxy, while the Square Kilometer Array Interferometer
would be able to image this D emission in the outer regions of nearby
galaxies; regions with low metal abundances.
These data should give the relationship between metal abundance
and D/H, and especially determine the fluctuations of D/H at a given metal
abundance which will
better determine Galactic chemical evolution, and, we expect,
allow an accurate prediction of primordial D/H independent of the QSO
observations.

\subsection{Solar System D/H}

The D/H in the ISM from which the solar system formed 4.6 Gyr ago can
be deduced from the D in the solar system today, since there should be no
change in D/H, except in the sun.

Measurement in the atmosphere of Jupiter will give the pre-solar D/H
provided (1) most of Jupiter's mass was accreted directly from the gas
phase, and not from icy planetessimals, which, like comets today, have
excess D/H by fractionation, and (2) the unknown mechanisms which deplete
He in Jupiter's atmosphere do not depend on mass.
Mahaffy et al. \cite{mah98} find D/H = $2.6 \pm 0.7$ \mf\ from the Galileo probe
mass spectrometer.
Feuchtgruber et al. \cite{feu99} used infrared spectra of the pure rotational
lines of HD at 37.7 $\mu$m to measure
D/H $=5.5^{+3.5}_{-1.5}$\mf\ in Uranus and
$6.5^{+2.5}_{-1.5}$\mf\ in Neptune, which are both sensibly higher
because these planets are known to be primarily composed of ices which
have excess D/H.

The pre-solar D/H can also be deduced indirectly from the present solar wind,
assuming that the pre-solar D was converted into \het .
The present \het / \hef\ ratio is measured and corrected for
(1) changes in \het /H and \hef /H because of burning in the sun,
(2) the changes in isotope ratios in the chromosphere and corona, and
(3) the \het\ present in the pre-solar gas.
Geiss \& Gloeckler \cite{gei98} reported D/H = $2.1 \pm 0.5$ \mf , later
revised to $1.94 \pm 0.36$ \mf\ \cite{glo99}.

The present ISM D/H $= 1.6 \pm 0.1$\mf\ is lower, as expected, and
consistent with Galactic chemical evolution models, which we now mention.

\subsection{Galactic Chemical Evolution of D}

Numerical models are constructed to follow the evolution of the abundances
of the elements in the ISM of our Galaxy.

The main parameters of the model include the yields of different stars,
the distribution of stellar masses, the star formation rate, and the
infall and outflow of gas. These parameters are adjusted to fit many
different data. These Galactic chemical evolution
models are especially useful to compare abundances at different epochs,
for example, D/H today, in the ISM when the solar system formed, and
primordially.

In an analysis of a variety of different models, Tosi et al. \cite{tos98}
concluded that the destruction of D in our Galaxy was at most a factor
of a few, consistent with low but not high primordial D.
They find that all models, which are consistent with all Galactic data,
destroy D in the ISM today by less than a factor of three.
Such chemical evolution
will destroy an insignificant amount of D when metal abundances are
as low as seen in the quasar absorbers.

Others have designed models which do destroy more D
\cite{van95}, \cite{timm97},
\cite{scul97}, \cite{oli99b}, for example, by cycling most gas through
low mass stars and removing the metals made by the
accompanying high mass stars from the Galaxy.
These models were designed to reduce high primordial D/H, expected
from the low \yp\ values prevalent at that time, to the low ISM values.
Tosi et al. \cite{tos98} describe the generic difficulties with these models.
To destroy 90\% of the D, 90\% of the gas must have been
processed in and ejected from stars. These stars would then release more metals
than are seen. If the gas is removed (e.g. expelled from the galaxy) to
hide the metals, then the ratio of the mass in gas to that in
remnants is would be lower than observed.
Infall of primordial gas does not help, because this brings in excess D.
These models also fail to deplete the D in quasar absorbers, because the
stars which deplete the D, by ejecting gas without D, also eject carbon.
The low abundance of carbon in the absorbers limits the destruction of
D to $<$1\% \cite{destruct of d}.

\subsection{Questions About D/H}

Here we review some common questions about D/H in quasar spectra.

\subsubsection{Why is saturation of absorption lines important?}

Wampler \cite{wam96} suggested 
that the low D/H values might be inaccurate because in some
cases the H absorption lines have zero flux in their cores; they
are saturated. Songaila, Wampler \& Cowie \cite{son nhi}
suggested that this well known problem might lead to errors in the
H column density, but later work, using better data and more
detailed analyses \cite{1937nhi} has shown that these concerns
were not significant, and that the initial result \cite{dh
details} was reliable.

Neutral deuterium (D~I) is detected in Lyman series absorption lines, which
are adjacent to the H~I lines. The separation of 82 \kms\ is easily resolved
in high resolution spectra, but it is not enough to move D out of the
absorption by the H. The Lyman series lines lie between 1216\AA\
and 912\AA , and can be observed from the ground at redshifts $>2.5$.

Ideally, many (in the best cases $>20$) Lyman lines are observed, to help
determine the column density (\nhi , measured in H~I atoms per \cmm\ along
the line of sight) and
velocity width (b values, $b = \sqrt{2} \sigma$,
measured in \kms ) of the H.
But in some cases only \Lya\ has been observed (Q1718+4807, APM 08279+5255),
and these give highly uncertain D/H, or no useful information.

The column densities of H and D are estimated from the precise shapes of their
absorption lines in the spectra. For H, the main difficulties are the accuracy
of the column density and the measurement of the distribution in
velocity of this H.
For D the main problem is contamination by H, which we discuss below.

It is well known that column densities are harder to measure when absorption
lines become saturated.
The amount of absorption increases linearly with the column density as long
as only a small fraction of the photons at the line central wavelength
are absorbed. Lines saturate when most photons are absorbed.
The amount of absorption then increases with the log of the column density.

Wampler \cite{wam96} has suggested that D/H values could be 3 -- 4 times
higher in Q1937--1009 than measured by Tytler, Fan \& Burles \cite{tyt96b}.
He argued that saturation of the H Lyman series lines could allow lower
\nhi . This would lead to residual flux in the Lyman continuum,
which would contradict the data, but Wampler suggested that the background
subtraction might have been faulty, which was not a known problem with HIRES.

Tytler \& Burles \cite{dh details} explained why Wampler's general concerns were not
applicable to the existing data on Q1937--1009.
Thirteen Lyman series lines were observed and
used to obtain the \nhi . The cross section for absorption (oscillator strength)
decreases by 2000 from the \Lya\ to the Ly-19 line.
This means that the lines vary significantly in shape, and this is readily
seen in spectra with high resolution and high signal to noise.
The background subtraction looked excellent because the line cores were near
zero flux, as expected.

Songaila, Wampler \& Cowie \cite{son nhi} measured the residual flux in the Lyman
continuum of the D/H absorber in Q1937--1009.
They found a lower \nhi\ and hence a higher D/H.
Burles \& Tytler \cite{1937nhi} presented a more detailed analysis of better
data, and found a lower \nhi , consistent with that obtained from the
fitting of Lyman series lines. 
They explained that Songaila,                    
Wampler and Cowie \cite{son nhi} had                                            
underestimated \nhi\ because they used poor estimates of the                    
continuum level and the flux in the Lyman continuum.

In summary, saturation does make the estimation of \nhi\ harder. Column
densities of H might be unreliable in data with low spectral resolution,
or low signal to noise, and when only a few Lyman lines are observed. The above
studies show that it is not a problem with the data available on
Q1937--1009, Q1009+2956, Q0014+8118 and Q0130-4021.
For the first two quasars, we obtain the same answer by two independent methods,
and for the last three the higher order Lyman lines are not saturated.

Saturation is avoided in absorbers with lower \nhi ,
but then the D lines are weaker, and
contamination by H lines becomes the dominant problem.

\subsubsection{Hidden Velocity Structure}

To obtain D/H we need to estimate the column densities of D and H.
Column densities depend on velocity distributions, and when lines are
saturated, it is hard to deduce these velocity distributions.
Similar line profiles are made when the velocity dispersion is increased to
compensate for a decrease in the column density.
We mentioned above that this degeneracy is broken when we observe lines
along the Lyman series. For Q1937--1009, which has the most
saturated H lines of the quasars under discussion, Burles \& Tytler \cite{bur98b}
showed that the D/H did not change for arbitrary velocity structures,
constrained only by the spectra.
The same conclusion was obtained for Q1009+2956 \cite{bur98a}.
The favorable results for these two quasars do not
mean that we will always be able to break the
degeneracy. That must be determined for each absorption system.

There are two reasons why hidden velocity structure is not expected to be a
major problem.
First, we are concerned about hidden components which have high
columns and low enough velocity dispersions that they hide inside
the wider lines from lower column gas.
Such gas would be seen in other lines which are not saturated:
the D lines and the metal lines from ions with similar (low) ionization.
Second, we search for D in absorbers with the simplest velocity
distributions. They tend to have both narrow overall velocity widths and low
temperatures, which makes it much harder to hide unseen components.
Typically, the main component accounts for all of the absorption in
the higher order Lyman lines, and these lines are too narrow for
significant hidden absorption.

\subsection{Correlated Velocity Structure: Mesoturbulence}

In a series of papers, Levshakov, Kegel \& Takahara \cite{lev1937}, 
\cite{lev1718}, \cite{mci}
have demonstrated a viable alternative model for the velocity distribution.

In most papers, absorption lines are modelled by Voigt profiles.
The line width is the sum of the
thermal broadening, turbulent broadening, and the instrumental resolution,
each of which is assumed to be Gaussian.
When an absorption line is more complex than a single Voigt, gas centered
at other velocities is added to the model. As the signal to noise
increases, we typically see that more velocity components are required to
fit the
absorption. Each component has its own physical parameters:
central velocity, velocity dispersion (rms of thermal and turbulent broadening),
ionization, column densities and elemental abundances.
Prior to its use with quasars, this fitting method was
developed for the ISM, where it represents gas in spatially separate clouds.

Levshakov and co-workers have proposed a different type of model,
the mesoturbulent model, in which
the gas velocities are correlated, and the column density per unit velocity
is varied to fit the absorption line profiles.
They assume that the absorption comes from a single region in space, and
they calculate the distribution of the gas density down the line of sight.
To simplify the calculations, in early Reverse Monte-Carlo models,
they assumed that the gas temperature and density were
constant along the line of sight, which is not appropriate
if there are separate discrete clouds of gas with differing physical conditions.

The effects of mesoturbulence on the D/H absorbers towards
Q1937--1009 \cite{lev1937}, Q1009+2956 \cite{1009} and
Q1718+4807 \cite{lev1718}
were examined in detail using this early model.
In the first paper they allowed the \nhi\ to vary far from
the observed value (\nhi $=7.27 \times 10^{17}$ \cite{1937nhi}),
and consequently they found a variety of \nhi , but when the \nhi\ is held
within range, the D/H is 3.3\mf , exactly the same as with the usual model
\cite{bur98b}.  For the second quasar, the
D/H obtained is again similar to that obtained in the usual way.
The results are the same as with the usual model
in part because the H and D line widths are dominated by thermal and
not turbulent motions, and for these two quasars the total \nhi\ is not
affected, because it
is measured from the Lyman continuum absorption, which does not depend on
velocity.

Recently they have developed a new model called MCI
\cite{08279}, \cite{mci}
appropriate for absorption systems which sample different densities.
They now use H~I and metal ions
to solve for two random fields which vary independently along the line
of sight: the gas density and the peculiar velocities.
This model allows the temperature, ionization and density to all
vary along the line of sight.

The mesoturbulent model of Levshakov et al. \cite{1009} and the
microturbulent Voigt model give the same column densities and other parameters
when one of the following conditions apply:
1) The line of sight through the absorbing gas traverses many
correlation lengths.
2) If each velocity in a spectrum corresponds to gas at a unique
spatial coordinate.
3) The absorbing regions are nearly homogeneous, with at most small
fluctuations in density or peculiar velocities, or equivalently,
thermal broadening larger than the turbulent broadening.

The Voigt model could give the wrong result when two or more
regions along the line of sight, with differing physical conditions,
give absorption at the same velocity.
A remarkable and unexpected example of this was reported by Kirkman \& Tytler
\cite{kirk99} who found a Lyman limit system which comprised five main velocity
components.
Each component showed both C~IV and O~VI absorption at about the same
velocity, but in each of the five components, the O~VI had a larger
velocity dispersion, and hence came from different gas than the C~IV.
While this LLS is much more complex than those in which we can see D, this
type of velocity structure could be common.

All authors other than Levshakov and collaborators
use standard Voigt fitting methods to determine
column densities, for several reasons.
The Voigt method was used, with no well known problems, for many decades
to analyze absorption in the ISM, and the ISM is well modeled by
discrete clouds separated in space.
The Levshakov et al. \cite{1009} methods are more complex.
In early implementations, Levshokov et al. \cite{1009} made assumptions which are
not suitable for all absorbers.
The current methods require weeks of computer time, and
in many cases the two methods have given the same results.

We conclude that, when we have sufficient data,
velocity structure is not a problem for the absorbers like those
now used for D/H.

\subsubsection{Was the primordial D high but depleted in the absorbers?}

The idea here is that the average BBN D/H was high, and it has been
depleted in the three absorbers which show low D.
There are two options: local depletion in some regions of the universe,
and uniformly global depletion.
We conclude that there is no known way to deplete D locally, and
global depletion seems unlikely.

First we list seven observations which together rule out local depletion,
including that suggested by Rugers \& Hogan \cite{ru96a}.

1. We note that D/H is also low in our Galaxy, and that
Galactic chemical evolution accounts for the difference from the low
primordial D. Hence we know of four places where D is low and
consistent with a single initial value.

2. If the BBN D/H was high, let us say ten times larger at 34\mf , then
the depletion in all four, widely separated in space, must be by a
similar factor:
Q1937--1009: $0.90 \pm 0.02$;
Q1009+2956: $0.88 \pm 0.02$;
Q0130-4021: $>0.80$;
local ISM in our Galaxy: 0.86 -- 0.93,
where for the Galaxy alone we assume that Galactic chemical evolution reduced
the initial D/H by a factor of 1.5 -- 3 \cite{tos98}.

3. The quasar absorption systems are large -- a few kpc along the line of sight
\cite{dh details}, far larger than can be influenced by a single
star or supernovae. The gas today in the local ISM is a mixture
of gas which was also distributed over a similar large volume prior to
Galaxy formation.

4. The abundance of the metals in the quasar cases are very low; too
low for significant ($>1$\%) destruction of D in stars \cite{destruct of d}.

5. The quasar absorbers are observed at high redshifts, when the universe is
too young for low mass stars ($<2$ solar masses) to have evolved to
a stage where they eject copious amounts of gas.

6. The quasar absorbers are observed at about the time when old stars
in the halo of our Galaxy were forming.
These stars may have formed out of gas like that seen in the quasar spectra, but
with high density.
We expect that much of the gas seen in absorption is in the outer
halo regions of young galaxies, and that some of it was later incorporated
into galaxies and halo stars.

7. The ratio of the abundances of Si/C in the quasar absorbers
is similar to that in old stars in the halo of our Galaxy.
This abundance ratio is understood as the result of normal chemical
evolution.

\bigskip
Global destruction of D prior to $z=3$, or
in the early universe, remains a possibility, but it seems contrived.

Gnedin \& Ostriker \cite{bh destruction of d} discuss photons from early black 
holes.  Sigl et al. \cite{d destruction by photons} show that this 
mechanism creates 10 times more \het\
than observed, and  Jedamzik \& Fuller \cite{destruct of d} find the
density of gamma ray sources is improbably high.

Holtmann, Kawasaki \& Moroi \cite{decay part a}, \cite{decay part b} showed that
particles which decay just after BBN might create photons which could
photodissociate D. With very particular parameters, the other nuclei are not
changed, and it is possible to get a D/H which is lower than from SBBN with the
same \ob . Hence low D and low \yp\ can be concordant.
An exception is \lisx\ which is produced with \lisx / H $\simeq 10^{-12}$,
which is about the level observed in two halo stars.
There is no conflict with the usual conclusion that most \lisx\ is made
by Galactic cosmic rays prior to star formation, because the
observed \lisx\ has been depleted by an uncertain amount.
This scenario has two difficulties: Burles (private communication) notes that
there would be a conflict with the \ob\
measured in other ways, and it seems unlikely that the hypothetical particle
has exactly the required parameters to change some abundances slightly,
within the range of measurement uncertainty, but not catastrophically.

Most conclude that there are no likely ways to destroy or make significant D.

\subsubsection{Could the D/H which we observe be too high?}

The answer to this question from Kirshner is, that the D/H could be
slightly lower than we measure, but not by a large amount.
We discuss two possibilities: measurement problems and biased sampling of the
universe.

First we consider whether the D/H in the quasar absorbers could be
less than observed.
This can readily happen if the D is contaminated by H, but a large reduction
in D/H is
unlikely because the D line widths match those expected in Q1937--1009 and
Q1009+2956.
We do not know how the ISM D/H values could be too high, and
Galactic chemical evolution requires primordial D/H to be larger than
that in the ISM, and similar to the low value from quasars.
Hence it is unlikely that the D/H is much below the observed value.

Second, we consider whether the absorbers seen in the quasar spectra are
representative.  The absorbers are biased in three ways: they represent regions
of the universe with well above (100 -- 1000 times) the
average gas density at $z=3$, and amongst such
high density regions, which are observed as Lyman Limit absorption
systems, they have relatively
low temperatures ($2 \times 10^4$ K), and simple quiescent velocity structures.
The last two factors are necessary to prevent the H absorption from
covering up that from D, while the high density follows from the
high density of neutral H which is needed to give detectable neutral D.
It is likely that the gas in the absorbers at $z=3$ has by today fallen
into a Galaxy, though this is not required because some gas will be heated
as galaxies form, preventing infall.
The low temperatures and quiescent velocities argue against violent
astrophysical events, and there are no reasons to think that the absorbers
are any less representative than, say, the gas which made up our Galaxy.

We should also consider whether the quasar absorbers might be unrepresentative
because of inhomogeneous BBN. In this scenario regions with above average
density will have below average D/H, but the evolution of density
fluctuations could be such that the low density regions
fill more volume \cite{inhom and high oba}, \cite{inhom and high obb}, so that they are more
likely to dominate the observed universe today. In that scenario the \ob\
derived from the D/H would be below the universal average, and the observed
(low) value of D/H would be ``high" compared to expectation for SBBN
with the same \ob .
This scenario will be tested when we have observations of many more quasars.

measurements of D/H towards QSOs. 

\subsubsection{Is there spatial variation in D/H towards quasars?}

It seems highly likely that the D is low in the three quasars which
show low D, and we discussed above why it is hard to imagine how this
D could have been depleted or created since BBN.
Hence we conclude that the low D/H is primordial.

Are there other places where D is high?
All quasar spectra are consistent with a single low D/H value.
The cases which are also consistent with high D are readily explained by
the expected H contamination. We now explain why we have enough data
to show that high D must be rare, if it occurs at all.

High D should be much easier to find than low D. Since we have not found
any examples which are as convincing as those of low D, high D must
be very rare. If D were ten times the low value,
the D line would be ten times stronger for a given \nhi , and could be
seen in spectra with ten times lower signal to noise, or 100 times
fewer photons recorded per \AA .
If such high D/H were common, it would have been seen many times in
the high resolution, but low signal to noise, spectra taken in the 1980's,
when the community was well aware of the importance of D/H.
High D would also have been seen frequently in the spectra
of about 100 quasars taken with the HIRES spectrograph on the Keck
telescope. In these spectra, which have relatively high signal to
noise, high D could be detected in
absorption systems which have 0.1 of the \nhi\ needed to detect
low D. Such absorbers are about 40 -- 60 times more common than
those needed to show low D/H, and hence we
should have found tens of excellent examples.

\subsubsection{Why is there lingering uncertainty over D?}

Today it is widely agreed that D is low towards a few quasars.
There remains uncertainty
over whether there are also cases of high D, for the following reasons:
\begin{itemize}
\item measurements have been made in few places;
\item contamination of D by H looks very similar to D, and resembles high D;
\item both the low \yp\ values reported during the last 25 years,
and the \lisv\ abundance in Spite plateau halo stars, with no correction for 
depletion, imply low \ob , low \ETA , and high D/H for SBBN; and
\item the first claims were for high D.
\end{itemize}

In most cases, the apparent conflicts over D/H values concern
whether the absorption near the expected position of D is mostly D or mostly H.
Steigman \cite{ste94a} and all observational papers discussed this
contamination of D by H.

Carswell et al. \cite{cars94} noted that contamination was likely in Q0014+813
and hence the D/H could be well below the upper limit.
Songalia et al. \cite{song94} stated:
``because in any single instance we can not rule out the possibility
of a chance H contamination at exactly the D offset, this result [the high D/H]
should be considered as an upper limit until further observations of other
systems are made."
Burles et al. \cite{bur99c} showed that Q0014+813 is strongly contaminated,
does not give a useful D/H limit. For Q1718+4807 we \cite{tyt98a} 
and Levshokov, Kegel \& Takahara \cite{lev1718} have
argued that contamination is again likely.

There are many reasons why contamination is extremely common:
\begin{itemize}
\item H absorption looks just like that from D,
\item H is 30,000 times more common,
\item spectra of about 50 quasars are needed to find one example of relatively
uncontaminated D,
\item high signal to noise spectra are needed to determine if
we are seeing H or D,
and
\item these spectra should cover all of the Lyman series and metal lines,
because we need all possible information.
\end{itemize}
When H contaminates D, the resulting D/H will be too high.

It is essential to distinguish between upper limits and measurements.
There are only two measurements (Q1937--1009 and Q1009+2956).
They are measurements because we
were able to show that the D absorption line has the expected width for D.
All other cases are upper limits, and there is no observational reason
why the D/H should be at the value of the limit.
In many cases, all of the D can
be H, and hence and D/H $= 0$ is an equally good conclusion from the data.

Only about 2\% of QSOs at $z \simeq 3$ have one
absorption systems simple enough to show D.
All the rest give no useful information on D/H.
Typically, they do not have enough H to show D, or there is no flux
left at the position of D. In such cases the spectra are
consistent with high, or very high, D/H, but
it is incorrect to conclude that D/H could be high in $\simeq 98$\% of
abosrption systems because these systems are not suitable to rule out
high D/H.
Rather, we should concentrate on the few systems which could rule out both
high and low D/H.

We will continue to find cases like Q1718+4807 which are consistent with
both low and high D/H.
As we examine more QSOs we will find some cases of contamination which
look exactly like D, even in the best spectra, by chance.
But by that time we will have enough data to understand the statistics of
contamination.
We will know the distribution function of the contaminating
columns and velocities, which we do not know today because the D/H absorbers
are a rare and special subset of all Lyman limit absorbers.
When absorbers are contaminated we
will find a different D/H in each case, because the
\nhi , velocity and width of the contaminating H are random
variables. But we will be able
to predict the frequency of seeing each type of contamination.
If there is a single primordial D/H then we should find many
quasars which all show this value, with a tail of others
showing apparently more
D/H, because of contamination. We will be able to predict this tail,
or alternatively, to correct individual D/H for the likely level of
contamination. When we attempted to correct for contamination in the past
\cite{tyt96b}, \cite{tb97}, \cite{tyt98a},
we used the statistics of H~I in the \lyaf\ because we do
not have equivalent data about the H~I near to the special LLS which
are simple enough to show D/H.
Such data will accumulate at about the same rate as do measurements of D/H,
since we can look for fake D which is shifted to the red (not blue) side
of the H~I.

There are large differences in the reliability and credibility of different
claimed measurements of D/H in quasar spectra, and hence much is missed
if all measurements are treated equally.
It also takes time for the community to criticize and absorb the new results.
Early claims of high D/H \cite{ru96a}, \cite{ru96b} 
in Q0014+8118 are still cited in a few recent papers, after later measurements 
\cite{bur99c} with better data, have shown that this quasar gives
no useful information, and that the high D/H came from a ``spike" in the data
which was unfortunately an artifact of the data reduction.

In summary, the lack of high quality
spectra, which complicates assessment of contamination by H,
is the main reasons why there remains uncertainty over whether
some absorbers contain high D.

\subsubsection{Why we believe that the D/H is Primordial}

Here we review why we believe that the low D/H is
primordial. These arguments are best made without reference to the other
nuclei made in BBN, because we wish to use the abundances of these nuclei
to test SBBN theory. 

\begin{itemize}
\item D/H is known to be low in four widely separated locations:
towards three quasars, and in the ISM of our Galaxy.
\item The extraction of D/H from quasar spectra is extremely direct,
except for corrections for contamination by H, which make D/H look too
large.
\item Since contamination is common, all data are consistent with low D/H,
and no data require high D/H.
\item High D/H is rare, or non-existent, because it should be easy to see
in many existing spectra, but we have no secure examples.
\item The low D/H in the quasars, pre-solar system and in the ISM today are
all consistent with Galactic chemical evolution.
\item The quasar absorption systems are large -- many kpc across, as was 
the initial volume of gas which collapsed to make our Galaxy.
\item The abundance of the metals in the quasar cases are very low,
and much too low for significant ($>1$\%) destruction of D in stars.
\item The quasar absorbers are observed at high redshifts, when the universe is
too young for low mass stars to have evolved to a stage where they
eject copious amounts of gas.
\item The ratio of the abundances of Si/C in the absorbers is normal for
old stars in the halo of our galaxy, indicating that these elements
were made in normal stars.
\item In the quasar absorbers, the temperatures and velocities are low, which
argues against violent events immediately prior to the absorption.
\item If BBN D/H were high, the hypothetical destruction of D would have to
reduce D/H by similar large amounts in all four places.
\item The above observations make local destruction of D unlikely.
\item There are no known processes which can make or destroy significant D.
\item Global destruction of D by photodissociation in the early universe
requires very specific properties for a hypothetical particle, and
is limited by other measures of \ob .
\end{itemize}

\subsubsection{Conclusions from D/H from quasars}

Most agree that D is providing the most accurate \ETA\ value
\cite{sch98a}, although some have one remaining
objection, that there might also be quasar absorbers which show
high values of D/H \cite{aud98}, \cite{oli99b}.

The D/H from our group (Burles \& Tytler \cite{1937nhi}, \cite{bur98a}, \cite{bur98b}),
together with over 50 years of theoretical work and laboratory measurements
of reaction rates, leads to the following values for cosmological parameters
(unlike most errors quoted in this review, which are the usual
$1\sigma $ values, the following are quoted with 95\% confidence intervals):
\begin{itemize}
\item D/H = $3.4 \pm 0.5 \times 10^{-5} $ (measured in quasar spectra)
\item \ETA\ $= 5.1 \pm 0.5 \times 10^{-10} $ (from BBN and D/H)
\item $Y_p = 0.246 \pm 0.0014$ (from BBN and D/H)
\item $\rm ^7Li/H = 3.5 ^{+1.1}_{-0.9} \times 10^{-10}$ (from BBN and D/H)
\item 411 photons cm$^{-3}$ (from the CMB temperature)
\item $\rho _b = 3.6 \pm 0.4 \times 10^{-31}\rm gcm^{-3}$ (from CMB and \ETA\ )
\item $\Omega _b h^2 = 0.019 \pm 0.0024$ (from the critical density $\rho _c$)
\item $N_{\nu } < 3.20$ (from BBN, D/H and \yp\ data).
\end{itemize}

If we accept that D/H is the most accurate measure of \ETA ,
then observations of the other elements have two main roles. First, they show
that the BBN framework is approximately correct.
Second, the differences between the observed and predicted primordial
abundances teach us about subsequent astrophysical processes.
Recent measurements of \hef\ \cite{hefdata} agree with the predictions.
It appears that some \lisv\ has been destroyed in halo stars \cite{li
accretion} , and \het\ is both created and destroyed in stars.

\section{ Helium }

The high abundance of \hef\ allows accurate measurements
in many locations.
However, \hef\ is also produced by stars, and since such high
accuracy is required, the primordial abundance is best measured in locations
with the least amounts of stellar production.
High accuracy is desired, since D/H predicts
\yp\ to within
0.0014 ($\delta$\yp /\yp = 0.006, 95\% confidence), which is well
beyond the typical accuracy of astronomical abundance determinations.
In the local ISM, the amount of \hef\ from stars is about
$Y=0.01 - 0.04$; much less than \yp , but ten times the desired accuracy
for \yp .

Helium has been seen in the intergalactic medium, where Carbon abundances are
$<0.01$ solar, and possibly zero in much of the volume.
Strong absorption is seen from the He~II \lya\ line at 304\AA\ in the
redshifted spectra of quasars \cite{he ii first}, however it is difficult
to obtain an abundance from these
measurements, because nearly all He is He~III which is unobservable, and
we do not know the ratio He~II/He to within an order of magnitude.
However, the strength of the He~II absorption does mean that there is abundant
He in the intergalactic gas \cite{wad99}, which has very low metal abundances, 
which is consistent with BBN, and probably not with a stellar origin for the
\hef .

The best estimates of the primordial abundance of He are from
ionized gas surrounding hot young stars (H~II regions) in small galaxies.
The two galaxies with the lowest abundances have
1/55 and 1/43 of the solar abundance.
The \hef\ and H abundances come from the strengths of the emission
lines which are excited by photons from near by hot stars.

Values for \yp\ from these extragalactic H~II regions
have been reported with small errors for more than 25 years, e.g.:
\begin{itemize}
\item \yp $=0.216 \pm 0.02$ \cite{he regress}
\item \yp $=0.230 \pm 0.004$ \cite{leq79}
\item \yp $=0.234 \pm 0.008$ \cite{kun83}
\item \yp $=0.236 \pm 0.005$ \cite{pag86}
\item \yp $=0.228 \pm 0.005$ \cite{pag92}
\item \yp $=0.234 \pm 0.002 \pm 0.005$ \cite{osw99 yp}.
(random, and systematic errors)
\item \yp $=0.246 \pm 0.0014$ (95\% prediction from low D/H and SBBN).
\end{itemize}

These values are lower than the value now predicted by low
quasar D/H and they appear incompatible, because of the small errors. However
Skillman et al. \cite{he more uncertain} argued that errors could be much larger than quoted,
allowing \yp $ < 0.252$, and Pagel \cite{pag95} (and private communication 1994)
agreed this was possible.

The measurement of \yp\ involves three steps.
Emission line flux ratios must be measured to high accuracy, which requires
good detector linearity and flux calibration, and
corrections for reddening and stellar He~I absorption.
These fluxes must be converted to an abundance, which requires correction
for collisional ionization and neutral He.
Correction for unseen neutral He depends on the spectral senegy
distribution adopted for the ionizing radiation and might change \yp\ by
1 -- 2 percent.
Then the primordial abundance must be deduced from the $Y$ values in different
galaxies.

Izotov, Thuan \& Lipovetsky \cite{he data 94}, \cite{he data 97}
have been pursuing a major observational program to improve the determination
of \yp . They have found many more low metallicity galaxies and
have been reporting consistently higher \yp\ values,
most recently in their clear and persuasive paper \cite{hefdata}:
\begin{itemize}
\item \yp $=0.244 \pm 0.002$ from regression with O/H and
\item \yp $=0.245 \pm 0.001$ from regression with N/H.
\end{itemize}

The four main reasons why these values are higher are as follows, in order of
importance \cite{hefdata}, \cite{izw18}, \cite{he systematics},
(Skillman, and Thuan personal communication 1998).

1. When stellar He~I absorption lines underlying He emission lines are not
recognized, the derived \yp\ is too low. This is a important for IZw18
\cite{izw18} which has the lowest metallicity and hence great
weight in the derivation of \yp , and perhaps for many other galaxies.

2. The emission line fluxes must be corrected
for collisional excitation from the metastable level. At low abundances,
which correlate with high temperatures, these corrections can be
several percent.
The amount of correction depends on the density. There are no
robust ways to measure these densities, and differing methods, used by
different groups, give systematically different results. Izotov and Thuan
\cite{izw18} solve for the He~II density, while Olive, Skillman and Steigman \cite{osw99 yp}
use an electron density from the S~II lines.

3. Izotov \& Thuan \cite{izw18} have spectra which show weaker lines, and they use the
five brightest He lines, while Olive et al. \cite{osw99 yp} usually use only HeI 6678.

4. Izotov \& Thuan \cite{hefdata}, \cite{izw18} correct for fluorescent 
enhancement, which increases the $Y$ values from for a few galaxies.

For these reasons Izotov \& Thuan \cite{izw18} obtain higher $Y$ values
for individual
galaxies which have also been observed by Olive, Skillman \& Steigman \cite{osw99 yp},
and Izotov \& Thuan \cite{izw18} find a shallower slope for the regression to zero metal
abundance (see \cite{sch98a} Fig 6).
And most importantly, using higher quality Keck telescope spectra, they
obtain high \yp = $0.2452 \pm
0.0015$ (random errors), from the two galaxies with the lowest metal
abundances \cite{izw 18 SBS 0335}.

These measurement difficulties, combined with the recent improvements,
lead most to conclude that the \yp\ is in accord with the SBBN.
The Izotov \& Thuan \cite{hefdata} values are very close to the low D/H predictions,
while the lower \yp\ quoted by Olive \cite{oli99b}, $0.238 \pm 0.002 \pm 0.005$, is
also consistent when the systematic error is used.

It is clear that the systematic errors associated with the \yp\ estimates
have often been
underestimated in the past, and we propose that this is still the case,
since two methods of analyzing the same Helium line fluxes give results which
differ by more than the quoted systematic errors.
While the Izotov \& Thuan \cite{izw18} method has advantages, we do not know why
the method used by Olive, Skillman \& Steigman \cite{osw99 yp} should give incorrect answers.
Hence the systematic error should be larger than the differences
in the results: 0.007 using the most recent values, or 0.011 using earlier
results.

\section{\het\ }

The primordial abundance of \het\ has not been measured.
This is most unfortunate, since it is nearly as sensitive as
D to the baryon density during BBN.
\het\ is harder to measure than D because the difference in wavelength
of \het\ and \hef\ lines is smaller than for D,
and the Lyman series lines of He~II, main absorption lines of He in the
IGM, are in the far ultraviolet at 228 -- 304\AA\ which is hard to
observe because of absorption in the Lyman continuum of H~I at $< 912$\AA .

Rood, Steigman \& Tinsley \cite{he3 made} argued that it was unlikely that
\het\ could be used to supplement cosmological information from D
because low mass stars should make a lot of \het , increasing the current
ISM value to well above that in the pre-solar system ISM, and in
potential conflict with observations at that time.
This conflict has been confirmed.
Measurements do show enhanced \het\ in
Planetary nebulae, as expected from the production in the associated low
mass stars, but this is not reflected in the ISM as a whole.
The pre-solar and current \het\
abundances are similar \cite{het limits}, in contradiction with
expectation \cite{het prod}, \cite{het problem}, for unknown reasons.

It was suggested (\cite{d+he3 proposal}, see review by Hata et al. \cite{dh predict})
that the uncertainty over the amount of destruction of D
could be circumvented using the sum of
the abundances of D + \het , since the destroyed D should become \het ,
and \het\ is relatively hard to destroy.
The primordial D + \het\ should then be $\leq $ the same sum observed today,
as more \het\ is made in stars over time.
However, there are two problems with this scenario. First,
the \het\ should increase over time, which
it does not, implying that some stars destroy \het , and second,
the \het\ abundance
should be about constant in the ISM today, which it appeared not to be in
early data \cite{he3 varies}.
Hence, just prior to the measurement of D in quasars, most concluded that
D + \het\ in the Galaxy does not provide secure cosmological information
\cite{dh predict}, and summaries by \cite{yp regress}, \cite{sch98}.

Balser et al. \cite{bal99} report on a 14 year program to measure \het\ in the
Galactic H~II regions.
Using models for the gas density structure, they find an average
\het /H $= 1.6 \pm 0.5$ \mf\ for a sub-sample of seven simple nebulae.
No variation is seen with Oxygen abundance
over a factor of ten, and there is little scatter \cite{he3 metals}.
This value may represent the average in the ISM today, but it
is not known how to use this to obtain primordial abundances.

These measurements are relevant to stellar nucleosynthesis
and Galactic chemical evolution, and are consistent with a
cosmological origin for the
\het , but we suggest that gas with much lower metal abundances will need to
be observed to derive a secure primordial abundance for \het .

\section{Lithium}

Lithium is observed in the solar system, the atmospheres of a wide variety of
stars and in the ISM.
Arnould \& Forestini \cite{arnould89} review light nuclei abundances in a variety of
stars and related stellar and interstellar processes, while
halo stars are reviewed by
\cite{li sys}, \cite{van99},
\cite{li accretion} and \cite{li summary}.

Old halo stars which formed from gas which had low iron abundances
show approximately constant \lisv $/H \simeq 1.6 \times 10^{-10}$ and
little variation with
iron abundance or surface temperature from 5600 -- 6300 K. The lack of variation amongst
these ``Spite plateau" stars \cite{early lisv}
(references in \cite{oli99b}) shows that their \lisv\ is close to primordial.

Since the halo stars formed
about ten times more \lisv\ has been produced in the inner Galaxy.
Abundances of \lisv /H $ \simeq 10^{-9}$ are common, although some stars
show more, presumably because they make \lisv .
Stars typically destroy \lisv\ when they evolve, accounting for the
low abundances, $<10^{-11}$, in evolved stars. Stars with
deeper convection zones, such as halo stars with lower surface temperatures,
show less \lisv , because they have burnt it in their interiors.

Here, and in the next section on \lisx , we will the following topics:

\begin{itemize}
\item measurement of current surface abundances on the Spite plateau,
\item change in \lisv\ with iron abundance,
\item creation of \lisv\ and \lisx\ after BBN and prior to halo star formation,
\item depletion of these nuclei in the halo stars,
\item stars with differing \lisv , and
\item gravitational settling.
\end{itemize}

The recent homogeneous data on 22 halo stars with a narrow range of
temperature on the ``Spite plateau"
have very small random errors and show that most (not all) stars with similar
surface properties have the same \lisv /H \cite{li accretion}.
Earlier data showed more scatter, which some
considered real (references in \cite{lisx BD}), and hence evidence
of depletion.

The Ryan, Norris, \& Beers \cite{li accretion} sample shows a clear increase of
\lisv\ with iron abundance, as had been found earlier.
This trend appears to be real, because the data and stellar atmosphere
models used to derive the \lisv\ abundance do not depend on metallicity. But
it was not found by Bonifacio \& Molaro
\cite{lith ref}, perhaps because of larger scatter in temperatures and iron abundance.
This trend is not understood, and there are several possible explanations.
It may have been established in the gas from which the stars formed, perhaps
from cosmic rays in the ISM, or from AGB stars. Alternatively, we
speculate that it might instead relate to depletion of the \lisv  in the stars.
In either case, the BBN \lisv\ will be different from
that observed: smaller if the \lisv\ was created prior to the star formation,
and higher if the trend is connected to destruction in the stars.
More on this below.

Creation of \lisv\ in the ISM by cosmic ray spallation prior to the
formation of the
halo stars is limited to 10 -- 20\% because Be would also be enhanced by
this process \cite{li accretion}, \cite{oli99b}.

A clear summary of arguments for and against significant
depletion is given by  Cayrel \cite{li sys}.
There are two main reasons why depletion is believed to be small:
the negligible dispersion in \lisv\ for most halo stars on the plateau,
and the presence of \lisx .
The main arguments for depletion are that it is expected, it clearly
occurs in some stars, some halo stars on the plateau
show differing abundances, and star in the globular cluster M92
which should have similar ages, composition and structure,
show a factor of two range in \lisv .

Different depletion mechanisms include
mixing induced by rotation or gravity waves, mass loss in stellar winds
and gravitational settling.
Some models predict either variation from star to star,
or trends with temperature, which are not seen for the stars on the plateau.
For example, the rotationally induced mixing model
implies that stars with different angular momentum histories
will today show different \lisv .
Ryan, Norris \& Beers \cite{li accretion} find that the small scatter in their data,
especially after the removal of the correlation with the iron abundance,
limits the mean depletion in these models to $<30$\%, much
less than the factor of two needed to make \lisv\ agree exactly with the
predicted abundance from low D/H.

Some stars which should lie on the plateau have very low \lisv ,
while others show a range of abundances (see ref. in \cite{li accretion}).
Differences are also seen between halo field stars
\cite{li accretion} and stars in the globular cluster M92
\cite{M92 li}, \cite{M92 li var}, which show a factor of two
spread in \lisv . These observations are not understood.

Gravitational settling (diffusion) of heavier elements reduces the
\lisv\ in the atmospheres of stars.
However, the depletion should be most in the
hottest (highest mass) stars, which is not seen, and not understood.
Vauclair \& Charbonnel \cite{li deplete winds} proposed that small stellar winds might be
balancing the settling. Vauclair \& Charbonnel \cite{li deplete calc} noted that
the peak abundances inside the stars are independent of
both mass and iron abundance. Normal stellar models predict that
these peak abundances will not be seen in the stellar atmospheres,
because convection does not reach this far down into the stars.
However they point out that if some mechanism does
mix gas from the \lisv\ peak zone into the bottom of the convection zone
then the stars on the plateau would have similar abundances as observed.
Assuming that the observed abundances are those
from the peaks inside the stars, they find that the initial abundance in
the stars was \lisv /H $= 2.2 \pm 0.6 \times 10^{-10}$, without free parameters,
which is still below but statistically consistent with the prediction from
low D of $3.5 _{-0.9}^{+1.1}$ \mt .

\subsection{Primordial \lisv }

Ryan, Norris and Beers \cite{li accretion} conclude \lisv /H $\simeq 10^{-10}$, with
small random errors and three
sources of systematic error, each up to a factor of 1.3,
from the effective temperatures, stellar atmospheres and enhancement prior
to star formation.
Bonifacio \& Molaro \cite{lith ref} found \lisv /H $= 1.73 \pm 0.05 \pm 0.2$ \mt .
These abundances are both below the value of
$3.5 ^{+1.1}_{-0.9} \times 10^{-10}$ (95\%)  from BBN and our D/H, but
unlike \cite{li accretion}, we feel they are not inconsistent given
the quoted systematic errors, the lack of understanding of depletion,
and the variation amongst similar stars. We do not know how to estimate the
systematic errors connected with these issues.
Given the comparative simplicity of D/H, we prefer to use it and
SBBN, and we stick with our earlier suggestion
\cite{tyt96b} that
\lisv\ in the Spite plateau halo stars is depleted by about a factor of two.
Most, but not all agree that this is reasonable.
Depletion by much larger factors, which was discussed a few years back, is now
our of favor because of improved models. Improved modelling of rotational
mixing, has lead to better fits to high metal abundance (population I) stars,
which can be applied to halo (population II) stars, while the initial rotation
rates of the halo stars may be lower than was assumed (Deliyannis private
communication).

In summary, both the data and theory tells us that
the \lisv\ on the Spite plateau
is not exactly the primordial value. The correction is probably small,
less than a factor of two, but we do not yet know its value.

If we are to attain a primordial $^7$Li abundance we must
either (1) understand why its abundance varies from star to star, and learn to
make quantitative predictions of the level of depletion, or (2) make
measurements in relatively unprocessed gas.

We are optimistic that primordial \lisv\ will be measured to high precision.
Compared to D and He, the observations are simple:
15 -- 20  m\AA\ absorption lines in relatively empty spectra of often bright
stars (V=11). The best data have small errors.
We anticipate that further studies
will determine the amount of \lisv\ produced prior to the
formation of the stars, and the subsequent depletion in these stars.
The possible increase in \lisv\ with iron abundance is a clue, as are the
\lisx , Be and B abundances in the same stars.

\subsection{\lisx\ }

The primordial \lisx\ abundance has not been observed, but \lisx /H
has been measured in two stars on the Spite plateau. The abundance is well
below that expected from SBBN, but \lisv\ is used to help determine the
primordial BBN \lisx\ abundance in two ways.
First, the presence of \lisx\ limits the amount of destruction of \lisv ,
because \lisx\ is more fragile than \lisv .
Second, if the observed \lisx\ was made prior to the formation of the stars,
then some, perhaps much \lisv\ \cite{lisx BD},
may have been made by similar processes.
The first point is often presented as evidence that the \lisv\ on the Spite
plateau is close to primordial (e.g. less that a factor of two depletion,
according to \cite{d n 191b2b}), but the second point is cause for caution.

\lisx\ has been detected in only two stars on the Spite plateau, because the
absorption line at 6707.97\AA\ is weak and fully blended with
\lisv\ at 6707.81\AA . This is a difficult observation.
The \lisx\ makes the absorption line slightly asymmetric, and this
is detected using models of the line broadening, which are tested on
other absorption lines which are expected to have similar profiles because
they arise in the same layers of the stellar atmosphere.
Following the impressive
first detection by \cite{lisx first} and \cite{li6},
and Cayrel et al. \cite{lisx 3 stars} report 
\lisx /\lisv $ = 0.052 \pm 0.019$ in HD84937,
while Smith, Lambert \& Nissen \cite{lisx BD} report 
\lisx /\lisv $ = 0.06 \pm 0.03$ in BD+26 3578.
It is not known whether these detections are representative of halo stars on
the Spite plateau. Most assume that they are, but they could be above normal,
perhaps by a lot; Smith et al. report \lisx /\lisv $ = 0.00 \pm 0.03$ for six
other stars.

The SBBN makes \lisx /H $\simeq 10^{-13.9}$
\cite{be b li in sbbn}, \cite{lisx}, using the \ETA\ from D/H,
which is 500 times less than the measured abundance of
$7 \times 10^{-12}$ in the two halo stars.
The SBBN isotope ratio is \lisx /\lisv = 3 \mf ,
a factor of 2000 less than observed in these two stars.
This is not considered a contradiction with SBBN, because
\lisx , and some \lisv\ at the same time, can be made elsewhere.

The \lisx\ is usually assumed to have been present in the gas when the stars
formed, but it could be made later,
e.g. when cosmic rays strike the star or in stellar
flares \cite{M92 li}.
Production by cosmic rays in the ISM prior to the
star formation is most favored \cite{lisx a}.
With this assumption, the effects on \lisv\ can be calculated in two steps.
First, determine the ratio of \lisx / \lisv\ in the production process
(the production ratio).
Second, correct for the depletion of \lisx\ in the stars to determine the
initial abundance of \lisx .
The amount of \lisv\ produced along with the initial \lisx\ is then
specified.

Cosmic rays in the early ISM could have made \lisx\ and some \lisv\
prior to the formation of the Spite plateau halo stars. The production ratio
depends on the reaction and energies (e.g. \cite{lisx BD}).
Two reactions of cosmic rays in the ISM are considered to produce \lisx .
Smith, Lambert \& Nissen \cite{lisx BD} find that \lisx /Be ratios imply that
most \lisx\ was made in $\alpha -\alpha$ fusion reactions, rather than in
spallation (e.g. O + p $\rightarrow$ \lisx ) which is favored by
\cite{lisx a} and \cite{lisx}.
The production ratio is \lisx / --\lisv\ $\simeq 2$ for the
$\alpha -\alpha$ reaction.

Standard stellar models \cite{M92 li} predict that
much of the initial \lisx\ will have been destroyed
in the stars. The more that was destroyed, the more \lisx\ and non-BBN
\lisv\ should have been in the initial gas to give the observed abundances.
Depending on the destruction mechanism, the
destruction of \lisx\ may also destroy \lisv , but this is usually ignored.

When we choose the amount of depletion of \lisx , we fix the amount
present when the stars formed.
If the \lisx\ has been depleted by a large factor, $\simeq 100$, then the
stars would have begun with \lisx /\lisv\ similar to the production ratio,
and essentially all of the \lisv\ would be non-primordial \cite{lisx BD},
which is an unusual conclusion!

Ryan, Norris and Beers \cite{li accretion} assume that 50\% of the \lisx\ and none of
the \lisv\ was destroyed, and use a production ratio of 1.5 to conclude that
the BBN \lisv\ was 0.84 of that now in the stars.
Since nearly all observations of Li are made at low resolution,
the \lisx\ and \lisv\ lines are not resolved, they correct for the
\lisx\ . If the two stars with observed \lisx\ are normal,
then the BBN \lisv\ is about 79\% of the observed Li absorption.

Many other papers discuss this  topic. Olive \& Fields \cite{oli99a} give a summary.
Cayrel et al. \cite{lisx 3 stars} use models for the formation of Li, Be and B
and calculate the expected abundance of \lisx\ when the star formed, and
find that the observed abundance implies little depletion of \lisx , and a
\lisv\ depletion of less than 25\%.
Vangiono-Flam et al. \cite{lisx} also argue that \lisx\ is not much depleted,
and find that its BBN abundance, extrapolated back to before the
production by spallation, is compatible with a BBN abundance of
$3 \times 10^{-13}$ -- $5.6 \times 10^{-14}$.

All eagerly await the measurement of \lisx , together with Beryllium, in more
stars.

\section{Beryllium}

The primordial abundance of Beryllium has not been observed. The
production in SBBN is
$^9$Be/H $< 10^{-17}$ \cite{be b li in sbbn},
\cite{lisx},
orders of magnitude below the observed level. Inhomogeneous BBN
allows much higher abundances, possibly approaching detection
\cite{be b li in sbbn}.

Be is observed. It is created in the ISM when cosmic rays strike
C, N and O nuclei, and it is destroyed in stars.
It is difficult to use Be to constrain the cosmic ray production of Li because
the production ratio is highly model dependent \cite{li be ratio}.
Beryllium is observed in the atmospheres of halo stars, including those
on the Spite plateau.
Boesgaard et al. \cite{boe99} have found that Be increases with Iron, and that
Be increases 8 times faster than Oxygen, a rate consistent with
cosmic ray creation.
There is some evidence for a spread in Be as a given Fe/H, but no sign of
a primordial plateau, down to Be/H $= 10^{-13.5}$.

\section{Are the different nuclei concordant or is there a crisis?}

Nearly everyone believes that the primordial abundances are consistent with BBN
(e.g. \cite{ker96}, \cite{hefdata}, \cite{sch98a},
\cite{oli99b}, \cite{bur99c}, \cite{review inhom}),
but there are many lingering questions about the measurements.
The reader will readily detect the two attitudes described by Audouze
\cite{aud86}:
``optimistic", and ``agnostic and perhaps heretical" in many papers.
Each of us tends to adopt differing attitudes for each nucleus and
astrophysical processes. This review favors D/H because it is simple and
familiar.

Steigman \cite{ste94a} noted that there was ``a hint of an emerging crisis"
because the \hef\ abundances appeared to be lower than expected using the
\ETA\ from the other
nuclei, but he recommended much more careful study of the uncertainty in
BBN predictions, chemical evolution, and observational uncertainties
including systematic effects.
Hata et al. \cite{crisis} and Steigman \cite{ste96 conflict}
stated that ``there is a conflict", referring to the
differences in \ETA\ implied by low D and low \yp\ values.

Whether or not there is a crisis depends on the confidence assigned to
the answers to three questions:
\begin{itemize}
\item Is primordial D/H low everywhere, or are there also a some places with
high values?
\item Is \yp\ low, high, or uncertain?
\item Has \lisv\ in halo stars been depleted by a factor of two?
\end{itemize}

Some combinations of answers are not consistent with SBBN.
Recent data make low D/H seem secure in three quasars plus the ISM,
hence the issue is whether there are also other places with high primordial D.
Low D/H is compatible with high \yp\ and depleted \lisv , but not
with low \yp\ or undepleted \lisv .
High D is compatible with low \yp\ and undepleted \lisv , but it is
incompatible with the three sites which show low D/H and
with Galactic chemical evolution. A factor of ten D depletion would be
required in all four places.
Low \yp\ is compatible with undepleted \lisv\ and high D, but is incompatible
with the low D.

A good case has been made for high \yp , explanations
have been given why earlier results gave lower values, and the uncertainty
appears to be larger than quoted. Hence D and \yp\ are in agreement.

The \lisv\ observed in stars on the Spite plateau is lower than values
consistent
with low D. Depletion might provide an explanation, but the amount
of depletion and the dominant mechanism are not known. The lack of
scatter implies little depletion, less than expected,
which \cite{li accretion}, \cite{li summary} conclude is not sufficient to match low D.
Bonifacio \& Molaro \cite{lith ref} find a higher \lisv , but still below
the level required to match low D without depletion.

\section{Non-standard BBN}

The many different forms of non-standard BBN
have been reviewed by Coles \& Lucchin \cite{cos text} and Jedamzik \cite{non-sbbn review}.
Much work has been devoted to inhomogeneous baryon distributions during BBN,
additional relativistic particles, decaying particles,
large neutrino chemical potentials (e.g. \cite{lepton asym}),
sterile neutrinos (e.g. \cite{sterile yp}), magnetic
fields (e.g. \cite{prim mag fields}),
anti-matter domains (e.g. \cite{antimatter bbn}),
and alternative theories of gravity (e.g. \cite{scalar tensor grav}).

\subsection{Inhomogeneous BBN}

Following early discussion of inhomogeneous BBN (IBBN) by
Epstein \& Lattimer \cite{early inhom} and Hogan \cite{lg ob with inhom},
many detailed studies of
different types of inhomoegeneity have been published.  Malaney \& Mathews
\cite{predictions} and Kainulainen, Kurki-Suonio, \& Sihvola \cite{review inhom} give reviews.

IBBN has been discussed to allow larger \ob\ than
standard BBN, to allow differing values of D/H in the universe,
and to reconcile low \yp\ with low D/H values.

One exciting goal of this work was to determine whether inhomogeneity could
give the observed abundances with \ob\ much larger than the usual value,
and perhaps large enough to account for all gravitating matter, without
the need for non-baryonic dark matter
(e.g. \cite{lg ob with inhom}, \cite{bh destruction of d}, \cite{inhom and high oba}).
The best upper limit on \ob\ comes from the
lowest observed D/H, which until recently was in the ISM.
In standard BBN, a higher
\ob\ is ruled out because BBN would make less than the observed ISM
D/H, and no other way to make D is known.
In IBBN the D/H in the ISM comes from
low density regions, allowing a higher average density.
The current observations, with some exceptions, fit SBBN well, and hence IBBN
allows only a slight increase in \ob .

Inhomogenieties can be imagined over a wide range of distance scales.
The smallest scales, $< 10^{-5}$ pc, mix prior to BBN, leaving homogeneous SBBN.
Small scales mix during BBN. Intermediate scales which mix after BBN give
abundances which are constant in space today, but the abundances are different
from SBBN with the same \ob . Extra D would be made in regions with low density
during BBN, giving enhanced D/H everywhere today.
Large scales ($> 1$ kpc) may have avoided mixing, and could give different
D/H in different locations today. The near isotropy of the CMB limits
inhomogenieties to $< 1$ Mpc.

Jedamzik \& Fuller \cite{inhom and high obb}
found it difficult to match observed abundances of \lisv\ with large
scale primordial isocurvature baryon number fluctuations.
Most overly dense regions of the universe with
masses greater than the local baryon Jeans mass would have to
collapse (to prevent
observation of the \lisv\ which is
overproduced) and smaller scale fluctuations would have
to be absent or suppressed. Gnedin, Ostriker \& Rees \cite{bh bbn} and
Copi, Olive \& Schramm \cite{cos inhom} reached similar conclusions.
Copi, Olive \& Schramm
\cite{cos inhom97} also showed that large scale ($>> $1Mpc) isocurvature
perturbations conflict with the smoothness of the CMB, but do not
rule out inhomogeneity \cite{destruct of d}.

Kainulainen, Kurki-Suonio, \& Sihvola \cite{review inhom} review IBBN.
The \ob\ can be higher than in SBBN provided the distance scale of the
baryon inhomogeneity is near to optimal to maximize neutron diffusion effects.
The distance scale expected for inhomogeneities arising in
the electroweak transition are too small ($10^{-6}$ to $10^{-3}$pc today)
to have major effects, although
not below the accuracy of BBN abundance calculations. QCD inhomogeneities are
not so limited. However, a low D/H $< 5$ \mf\ still requires \yp $> 0.240$
even in IBBN, which helps reconcile low D/H and low \yp\ measurements,
especially when we accept that the errors on \yp\ are larger than quoted.

Rehm \& Jedamzik \cite{antimatter bbn} studied BBN in the presence of anti-matter domains.
Annihilation is preferentially on neutrons, and in a limiting case
the resulting universe is without light nuclei, in violation of the measured
abundances. With small amount of anti-matter, both the low \yp\ 
and low D/H measurements are matched.

Early IBBN results looked promising.
Today it appears that the scales are too small to have major effects, and
measurements of primordial abundances, especially upper limits on \lisv ,
with modest depletion ($<$ factor of two),
are usually used to give limits on the inhomogeneity,
rather than to argue that inhomogeneity helps explain discordant data or
allows different conclusions about \ob .

\subsection{The number of Relativistic Particles and their Decays}

The main idea here is that the \hef\ abundance depends on the
number of relativistic particles during BBN. Extra particles, such as
neutrinos or supersymmetric particles, which are relativistic
during BBN, lead to faster expansion, larger n/p and a larger \yp .

Steigman, Schramm \& Gunn \cite{ste77} calculated that BBN
limited the number of families to $N _\nu < 5$ to match the
\hef\ abundance. The range allowed by SBBN and laboratory measurements
have both narrowed over the years and agree well today
\cite{neutrino hist}, \cite{sch98a}.
A recent update \cite{bur99c} gives
$N _{\nu} < 3.20$ (95\%) from SBBN, although a larger range is obtained if
a wider variety of measured abundances are accepted \cite{lis99}.
March-Russell et al. \cite{lepton asym} note that additional relativistic degrees of
freedom are allowed if there is a large compensating asymmetry in the electron
neutrino number.

Shi, Fuller \& Abazajian \cite{sterile yp} follow the time evolution of a lepton
number asymmetry arising from active -- sterile neutrino transformations
during BBN. For $\nu _e $ mixing with $\nu _s$, \yp\ was allowed to change
from --1\% to +9\%, while $\nu _{\tau }$ or $\nu _{\mu }$ mixing with
$\nu _s$ allowed --2\% to +5\%. Hence the \yp\ predicted by low D/H in SBBN
could be lowered to 0.241, which is between the high and low
measurements.

For many years past, observations suggested that \yp\ was smaller
than expected for the low D/H in the ISM and now QSOs.
It is hard to make \yp\ lower, since this requires fewer, not more,
particles than in SBBN. Holtmann et al. \cite{decay part a}, \cite{decay part b}
proposed decays of neutrinos, but this is nearly ruled out by the Kamiokande
results on atmospheric neutrinos
\cite{review inhom}.

Lindley \cite{life of decay part} found that massive particles decaying into photons
must have lifetimes
in excess of a few thousand seconds, to avoid the destruction of BBN D.
Audouze, Lindley \& Silk \cite{decay part 1000 sec} noted that such radiative decays
could photodisintegrate
\hef\ and make D and \het , removing the upper bound on \ob .
Other references are given by Holtmann et al. \cite{decay part b}, who
discuss weakly interacting massive (100 GeV) particles which decay of
order $10^6$ s after BBN.
The authors give limits on the abundance and lifetimes of
gravitinos and neutralinos, for a wide range of light nuclei 
primordial abundances.

Kohri \& Yokoyama \cite{primordial bh} give limits on
the mass fraction in primordial black holes with masses
$10^8 - 3\times 10^{10}$ g which evaporate during BBN and change the abundances.

L{$\rm\acute o$}pez-Su{$\rm \acute a$}rez \& Canal \cite{inhom + decays} combine
inhomogeneous nucleosynthesis and
particles which decay at a late time to reassess the limits on \ob .
They find parameters which allow \ob $< 0.13 - 0.18 h_{70}^{-2}$
(\hsv\ is the Hubble constant in units of 70 \ho ). Such high \ob\ might
appear to remove the need for non-baryonic dark matter, but there would
then be conflicts with other measures of \ob , especially the baryon 
fraction in clusters of galaxies, if all those baryons were observable today.

\section{Cosmological Baryon Density}

The measurement of the baryon density is now a highly active area of
research.
In the coming years, we anticipate that higher accuracy measurements of the
baryon
density, from the CMB, clusters of galaxies, and the \lya\ forest, will
give a new rigorous test of BBN \cite{sch98a}.
This test can be viewed from two directions.
First, we can use the baryon density to fix the last free parameter in BBN,
and second, we can compare the different baryon density measurements, which
should be identical if SBBN is correct, and all baryons are counted in the
measurements made at later times.

In addition to BBN, the baryon density is measured in four ways:
in the IGM,
in clusters of galaxies,
using simulations of galaxy formation,
and directly from the CMB.
All agree
with the value from SBBN using low D/H, but today they are each about an
order of magnitude less accurate.

\subsection{\ob\ from the IGM Lyman-$\alpha$ forest absorption}

The gas in the IGM is observed through H~I \Lya\ absorption in the
spectra of all QSOs. Gunn \& Peterson \cite{gunn65} discussed how redshift
produces continuous
absorption in the ultraviolet spectra of QSOs.
Density fluctuations in the IGM trun this continuous absorption
into the \lyaf\
absorption lines. The IGM fills the volume of space, and at redshifts
$z>1$ \cite{where are baryons} it contains most of the baryons.

The baryon density is estimated from the total amount of H~I absorption,
correcting for density fluctuations which change the ionization.
The gas is photoionized, recombination
times are faster in the denser gas, and hence this gas shows more H~I
absorption per unit gas. Using the observed ionizing radiation from QSOs,
we have a lower limit on the ionizing flux, and hence a lower limit on the
ionization of the gas. If the gas is more ionized than this, then we
have underestimated the baryon density in the IGM.

Three different groups obtained similar results
\cite{ob from lyaf}, \cite{ob lyaf}, \cite{ob from lyaf zhang}:
\ob $> 0.035 h_{70}^{-2}$.
This seems to be a secure lower limit, but not if
the IGM is less ionized than assumed, because there is more
neutral gas in high density regions, and these were missing from
simulations which lack resolution.

We do not have similar measurements at lower redshifts, because the
space based data are not yet good enough, and the
universe has expanded sufficiently that simulations are either
too small in volume or lack resolution.
Cen \& Ostriker \cite{where are baryons} have shown that by today, structure formation may have
heated most local baryons to temperatures of $10^5-10^7~$K,
which are extremely hard to detect \cite{where are baryons},
\cite{halpha detect baryons}.

\subsection{Clusters of Galaxies}

Clusters of galaxies provide an estimate of the baryon density because
most of the gas which they contain is hot and hence visible.
The baryons in gas were heated up to 8 keV through fast collisions
as the clusters assembled.
The mass of gas in a cluster can be estimated from the observed X-ray
emission, or from the scattering of CMB photons
in the Sunyaev-Zel'dovich (SZ) effect. Other baryons
in stars, stellar remnants and cool gas contribute about 6\% to the
total baryon mass.

The cosmological baryon density is obtained from the ratio of the baryonic
mass to the total gravitating mass \cite{baryon fraction}.
Numerical simulations show that
the value of this ratio in the clusters will be similar to the cosmological
average, because the clusters are so large and massive, but slightly
smaller, because shock heating makes baryons more extended
than dark matter \cite{cluster baryon}, \cite{greece98a}.
The total mass of a cluster, $M_t$,
can be estimated from the velocity dispersion
of the galaxies, from the X-ray emission, or from the weak lensing of
background galaxies.
We then use \ob / \om $\simeq M_b/M_t$. The baryon fraction in clusters
in the last factor is about
$0.10 h_{70}^{-1}$ (SZ effect: \cite{sz baryon}), or
$0.05 - 0.13 h_{70}^{-3/2}$ (X-ray: \cite{cluster baryon fraction}), or
$0.11 h_{70}^{-3/2}$ (X-ray: \cite{cluster baryon q0}, \cite{arn99}).
Using \om $= 0.3 \pm 0.2$ from a variety of methods \cite{triangle},
we get \ob $\simeq 0.03$, with factor of two errors.
These \ob\ estimates are lower limits, since there might be
additional unobserved baryons.

\subsection{Local Dark Baryonic Matter}

The baryon density estimated in the \lyaf\ at $z \simeq 3$ and in local
clusters of galaxies are both similar to the that from SBBN using low D/H.
This implies that there is little dark baryonic matter in the universe
\cite{machos not baryonic}.
This result seems conceptually secure, since there
is little opportunity to remove baryons from the IGM at $z<3$ or
to hide them in dense objects without making stars which we would see
\cite{$<0.05$ ob remnants}, and
the clusters are believed to be representative of the contents of the universe
as a whole today. However, the numerical estimates involved are not
yet accurate enough to rule out a
significant density (e.g. 0.5 \ob ) of baryonic MACHOS.

\subsection{Simulations of the formation of Galaxies}

Ostriker (private communication) notes that the \ob\ can be constrained
to a factor of two of that derived from SBBN using low D/H by the
requirement that these baryons make galaxies.
Semi-analytic models can also address the distribution of baryons in
temperature and the total required to make observed structures
(Frenk \& Baugh, personal communication).

\subsection{CMB}

The baryon density can be obtained from the amplitude of the
fluctuations on the sky of the temperature of the CMB.
The baryons in the IGM at $z \simeq 1300$ scattered the CMB photons.
The amplitude of the fluctuations is a measure of \ob h$^2$, and other
parameters.
Published data favor large \ob , with large errors, however dramatic
improvements are imminent, and
future constraints may approach or exceed the accuracy of \ob\ from
SBBN \cite{kam96}, \cite{ob form cmb}.

\section{The Achievements of BBN}

Standard Big Bang Nucleosynthesis (SBBN)
is a major success because the theory is well understood,
close connections have developed between theory and observation, and
observations are becoming more reliable.

The early attempts to include physics in the mathematical model of the
expanding universe lead to an understanding of the creation of the elements
and the development of standard big bang theory, including the
predictions of the CMB.

The general success of SBBN is based on the robustness of the theory,
and the resulting predictions of the abundances of the light nuclei.
The abundances of \hef , \lisv\ and D
can be explained with a single value for the free parameter
\ETA , and the implied \ob\ agrees with other estimates.

This agreement is used to limit physics beyond that in SBBN,
including alternative theories of gravity, inhomogeneous baryon density,
extra particles which were relativistic during BBN, and decays of particles
after BBN. After decades owere f detailed study, no compelling major departures
from SBBN have been found, and few departures are allowed.

Using SBBN predictions and measured abundances,
we obtain the best estimates for the cosmological parameters
\ETA\ and \ob .

The abundances of D, \hef\ and \lisv\ have all been measured in gas where
there has been little stellar processing.
In all three cases, the observed abundance are near to the primordial
value remaining after SBBN. The D/H measured toward QSOs has the
advantage of simplicity: D is not made after BBN, there are no known
ways to destroy D in the QSO absorbers, and D/H can be extracted directly
from the ultraviolet spectra, without corrections.
There are now three cases of low D/H which seem secure. There remains
the possibility that D/H is high in other absorbers seen towards
other QSOs, but such high D must be very rare
because no secure cases have
been found, yet they should be an order of magnitude easier to find
than the examples which show low D.

We use low D/H as the best estimator of \ETA\ and the baryon density.
SBBN then gives predictions of the abundance of the other light nuclei.
These predictions suggest that \yp\ is high, as suggested by
Izotov, Thuan and collaborators. Low D also implies
that \lisv\ has been depleted by about a factor of two in the halo
stars on the Spite plateau, which is more than some expect.

The high \ob\ from SBBN plus low D/H is enough to account for
about 1/8th of the gravitating matter. Hence the remaining dark matter is not
baryonic, a result which was established decades ago using SBBN and D/H in
the ISM.

The near coincidence in the
mass densities of baryons and non-baryonic dark matter is perhaps explained
if the dark matter is a supersymmetric neutralino \cite{ellis susy dm ob}.

At redshifts $z \simeq 3$ the baryons are present and observed in
IGM with an abundance similar to \ob . Hence there was no dark, or missing
baryonic matter at that time. Today the same is true in clusters of galaxies.
Outside clusters the baryons are mostly unseen, and they may
be hard to observe if they have been heated to
$10^5$ -- $10^7$K by structure formation.

The number of free parameters in BBN has been decreasing over the years:
Fermi \& Terkovich gave nuclear reaction rates, the half-life of the
neutron was measured, and then the number of families of neutrinos was measured.
In standard BBN we are now left with one parameter, the baryon density,
which is today measured with D/H using SBBN. When, in the next few years,
this parameter is also measured, SBBN will have no free parameters.
When free parameters can be adjusted to obtain consistency
with the data, it is hard to tell if a hypothesis is correct.
The agreement between SBBN theory and measurement has grown stronger over the
decades, as more parameters were constrained by independent measurements, and
abundance measurements improved.
This is the most convincing evidence that BBN happened and has been understood.

This work was funded in part by grant G-NASA/NAG5-3237 and by NSF
grants AST-9420443 and AST-9900842. 
We are grateful to Steve Vogt, the PI for the Keck
HIRES instrument which enabled our work on D/H.
Scott Burles and Kim Nollet kindly provided the figures for this paper.
It is a pleasure to thank
Scott Burles,
Constantine Deliyannis,
Carlos Frenk,
George Fuller,
Yuri Izotov,
David Kirkman,
Hannu Kurki-Suonio,
Sergei Levshakov,
Keith Olive,
Jerry Ostriker,
Evan Skillman,
Gary Steigman
and Trinh Xuan Thuan for suggestions and many
helpful and enjoyable discussions.
We thank the organizers of this meeting, Lars Bergstrom,
Per Carlson \& Claes Fransson for their gracious hospitality.

\clearpage

\begin{figure}
\centerline{\psfig{file=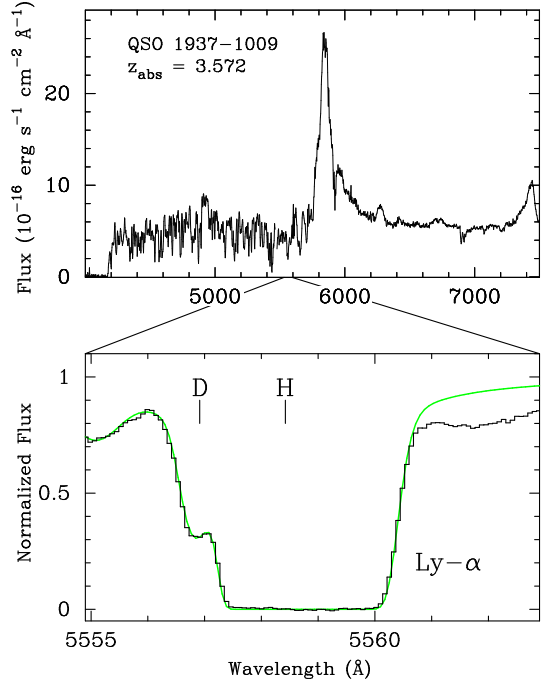, height=4.0in}}
\caption{\footnotesize{Optical spectrum of quasar 1937--1009, which
shows the best example of primordial D/H.
The top spectrum, from the Kast spectrograph on the 3-m telescope at
Lick observatory, is of low spectral resolution, and high signal to noise.
The continuum emission, from the accretion disk surrounding the black hole
at the center of the quasar, is at about 6 flux units.
The emission lines showing more flux (near 4950, 5820, 5940, 6230, 6700 \&
7420 \AA ) arise in gas near the quasar.
The absorptoin lines, showing less flux, nearly all arise in gas which is
well separated from, and unrelated to the quasar.
The numerous absorption lines
at 4200 -- 5800 \AA\ are H~I Ly$\alpha$ from the gas in the intergalactic
medium. This region of the spectrun is called the Ly$\alpha$ forest.
This gas fills the volume of the intergalactic medium, and the absorption lines
arise from small, factor of a few, fluctuations in the density of the gas
on scales of a few hundred kpc. The Ly$\alpha$ lines were all created by
absorption of photons with wavelengths of 1216\AA .
They appear at a range of observed wavelengths because they have different
redshifts. Hence Ly$\alpha$ absorption at 5800\AA\ is near
the QSO, while that at
5000\AA\ is nearer to us.
The abrupt drop in flux at 4180 \AA\ is caused by H~I Lyman continuum
absorption in the absorber at $z=3.572$. Photons now at $<4180$ \AA\ had
more than 13.6 eV when they passed though the absorber,
and they ionized its H~I.
The 1\% residual flux in this Lyman continuum region has been measured in
spectra of higher signal to noise (Burles \& Tytler 1997) and gives
the H~I column density, expressed as
H~I atoms per cm$^{-2}$ through the absorbing gas.
The lower plot shows a portion of a spectrum with much higher resolution
taken with the HIRES spectrograph on the Keck-1 telescope.
We mark the \lya\ absorption lines of H~I and D from the same gas.
The column density of D is measured from this spectrum.
Dividing these two column densities we find
D/H $= 3.3 \pm 0.3 \times 10^{-5}$ (95\% confidence), which is
believed to be the primoridal value, and using SBBN predictions, this
gives the most accurate measurements of $\eta$ and $\Omega_b$.}}
\end{figure}

\end{document}